\documentclass[]{spie}

\usepackage[]{graphicx}
\usepackage{amsmath, amssymb, amsfonts}
\usepackage[english]{babel}
\usepackage[]{epstopdf}
\usepackage{booktabs,caption,fixltx2e}
\usepackage[flushleft]{threeparttable}
\usepackage[font={small,it}, margin = 0.5cm]{caption}

\usepackage{wrapfig}
\usepackage{bigints}

\usepackage{hyperref}

\usepackage{subcaption}

\title{Fundamental limits to high-contrast wavefront control} 

\author{Johan Mazoyer\supit{a,b} \& Laurent Pueyo\supit{a}
\skiplinehalf
\supit{a} Space telescope Science Institute, 3700 San Martin Drive, Baltimore, MD 21218, USA\\
\supit{b} Department of Physics and Astronomy, Johns Hopkins University, Baltimore, MD, USA\\
}

\authorinfo{Further author information: contact Johan Mazoyer at jmazoyer@jhu.edu}
 
\begin{document} 

\maketitle 

\begin{abstract}
The current generation of ground-based coronagraphic instruments uses deformable mirrors to correct for phase errors and to improve contrast levels at small angular separations. Improving these techniques, several space and ground based instruments are currently developed using two deformable mirrors to correct for both phase and amplitude errors. However, as wavefront control techniques improve, more complex telescope pupil geometries (support structures, segmentation) will soon be a limiting factor for these next generation coronagraphic instruments.

In this paper we discuss fundamental limits associated with wavefront control with deformable mirrors in high contrast coronagraph. We start with an analytic prescription of wavefront errors, along with their wavelength dependence, and propagate them through coronagraph models. We then consider a few wavefront control architectures, number of deformable mirrors and their placement in the optical train of the instrument, and algorithms that can be used to cancel the starlight scattered by these wavefront errors over a finite bandpass. For each configuration we derive the residual contrast as a function of bandwidth and of the properties of the incoming wavefront. 

This result has consequences when setting the wavefront requirements, along with the wavefront control architecture of future high contrast instrument both from the ground and from space. In particular we show that these limits can severely affect the effective Outer Working Angle that can be achieved by a given coronagraph instrument.

\end{abstract}



\section{Introduction}
\label{sec:intro}

The current generation of high-contrast coronagraphic instruments on ground-based telescopes was designed to reach contrast levels of $10^{−6}$. These instruments \cite{macintosh08, beuzit08, martinache10,oppenheimer12} already rely heavily on high order deformable mirrors (DM) to correct for the effect of the atmosphere. 
The next generation of instrument will have to reach a $10^{−8}$ contrast level to observe Jupiter-like planets from the ground ELTS \cite{macintosh06,kasper08,davies10,quanz15c}, or a $10^{−10}$ limit for earth-like planets. These instruments will probably use at least two high order DMs to correct not only for the effect of the atmosphere, but also for the aberrations introduced by the optics, using state-of-the-art wavefront sensing\cite{borde06,paul13,riggs16,ndiaye16b, mazoyer14} and correcting techniques. In addition, as the size of the primary of these telescopes grows, so is their complexity. Both ground- and space-based telescopes will have to deal with large struts (e.g. WFIRST) and/or heavily segmented primaries (e.g. ELTs). These effects can be corrected using two DMs. The active correction for aperture discontinuities technique (ACAD)\cite{pueyo13,mazoyer15,mazoyer16b} was developed for this reason, and improved recently into the active correction for aperture discontinuities technique-opimized stroke minimization (ACAD-OSM, see paper \#10400-16, Mazoyer et al., in these proceedings) technique \cite{mazoyer16c,mazoyer17a,mazoyer17b}. However, the use of DMs for active correction comes with limitations that have to be understand to help the design of the instruments. 

\begin{figure}
\begin{center}
 \includegraphics[width = .48\textwidth]{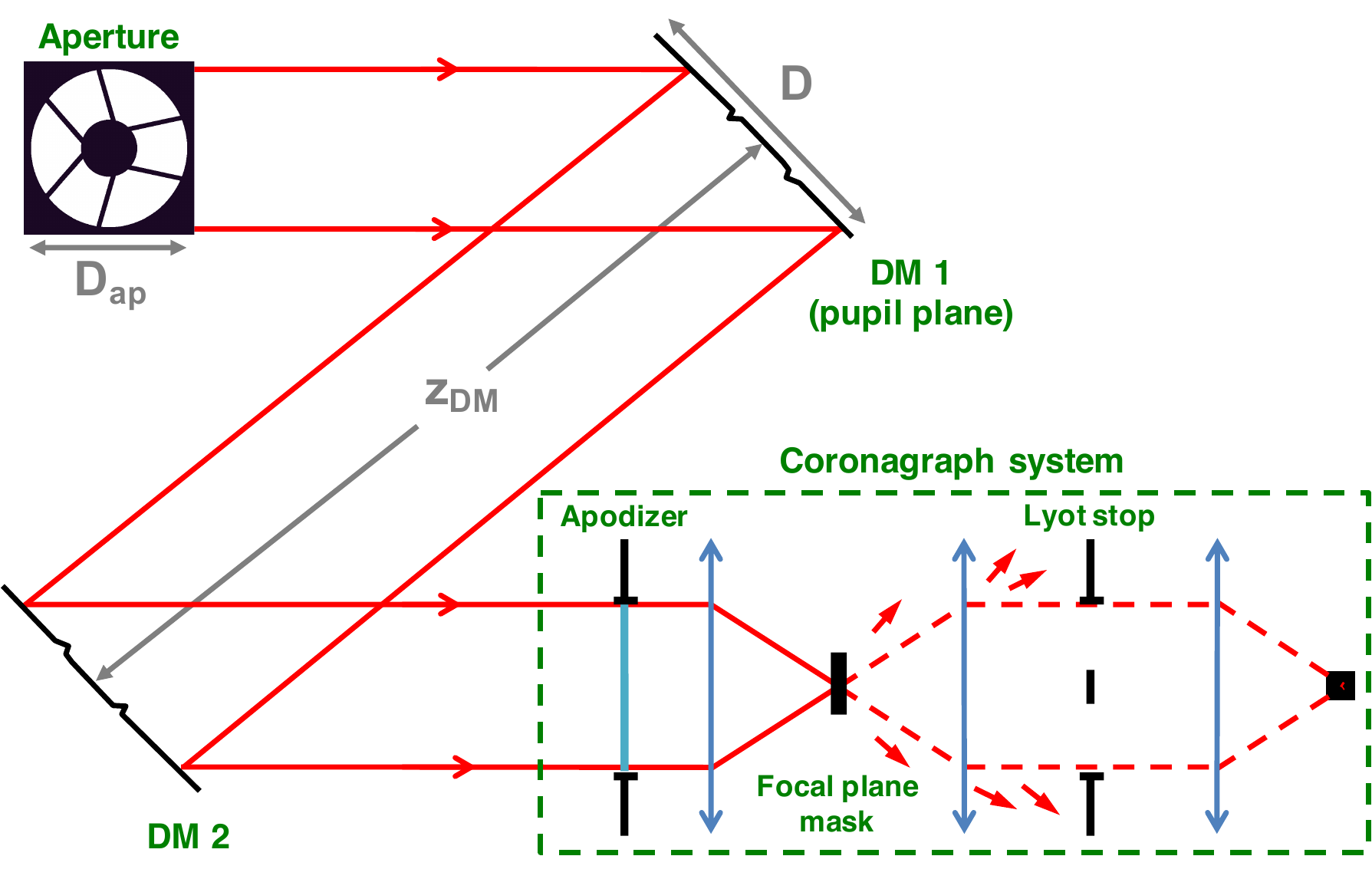}
 \end{center}
\caption[plop]
{\label{fig:optical_schema_2DM} Schematic representation of a two DM system and a coronagraph. We show the distances $z_{DM}$, D and $D_{ap}$ on this optical layout.}
\end{figure}

In this proceedings, we present some of the tools we developed to understand the intrinsic limitations of these active systems. These theoretical tools do not replace the complete end-to-end simulation that will be necessary to simulated and built these instrument, but they can help to predict and understand their results. We use the system showed in Fig.~\ref{fig:optical_schema_2DM}: a two DM system, followed by a coronagraph system. No coronagraph will not be studied in this particular. We will study both the case with phase and amplitude aberrations (discontinuities in the aperture) in entrance of the system. 

The two DMs is are square of size $D \times D$. The first DM is in pupil plane, as is the circular aperture, of diameter $D_{ap}$ (see Fig.~\ref{fig:optical_schema_2DM}). We slightly oversize the DMs compared to the pupil:
\begin{equation}
\label{eq:alphadef}
D = (1+\alpha)D_{ap}\,\,\,\,\,\,\,\,\, \alpha \ge 0
\end{equation}
In practice, we express $\alpha$ as a percentage (from 0\% to 30\%). The second DM is located at a distance $z_{DM}$ from the first DM. Due to the Fresnel propagation, the diameter of the beam expands between the first and second DM. We define the Fresnel number:
\begin{equation}
\label{eq:fnum}
F  = \dfrac{D^2}{\lambda z}.
\end{equation}
We show here that this parameter is the only relevant propagation parameter and in this whole article we place ourselves in the Fresnel number space. In particular, we define the Fresnel number of the DM setup at the central wavelength $\lambda_0$:
\begin{equation}
\label{eq:fnumDM}
F_{DM} = \dfrac{D^2}{\lambda_0 z_{DM}}
\end{equation}

In Section~\ref{sec:vignetting}, we analyze the phenomenon of vignetting by the second DM and shows that for a given Fresnel number of the system $F_{DM}$, this effect puts a limit on the OWA of the system. In Section~\ref{sec:talbot}, we analyze the limitation in terms of correction of these active system, first analytically and then using an end-to-end simulation to understand the observed effects. Finally, in the last section, we use the same formalism to show that the requirement on the optics quality upfront of the first DM in pupil plane puts a limit on the OWA of the system. This effect was already studied by Shaklan \& Green 2006 \cite{shaklan06}, and we generalize it to all systems.

\section{Vignetting due to the second DM}
\label{sec:vignetting}

We analyze the effect of the Fresnel number on the throughput of an off-axis planet located at an angular separation of $N\lambda_0/D_{ap}$ from the on-axis star. We first use a geometrical argument to derive an theoretical law linking the off-axis transmission of the system with the Fresnel number. Then, we use numerical simulations to show the precise impact of this parameter on the performance in off-axis throughput.

\subsection{Geometric analysis of the effect of vignetting}
\begin{figure}
\begin{center}
 \includegraphics[trim= 0.7cm 0cm 0cm 0.5cm, clip = true,width = .48\textwidth]{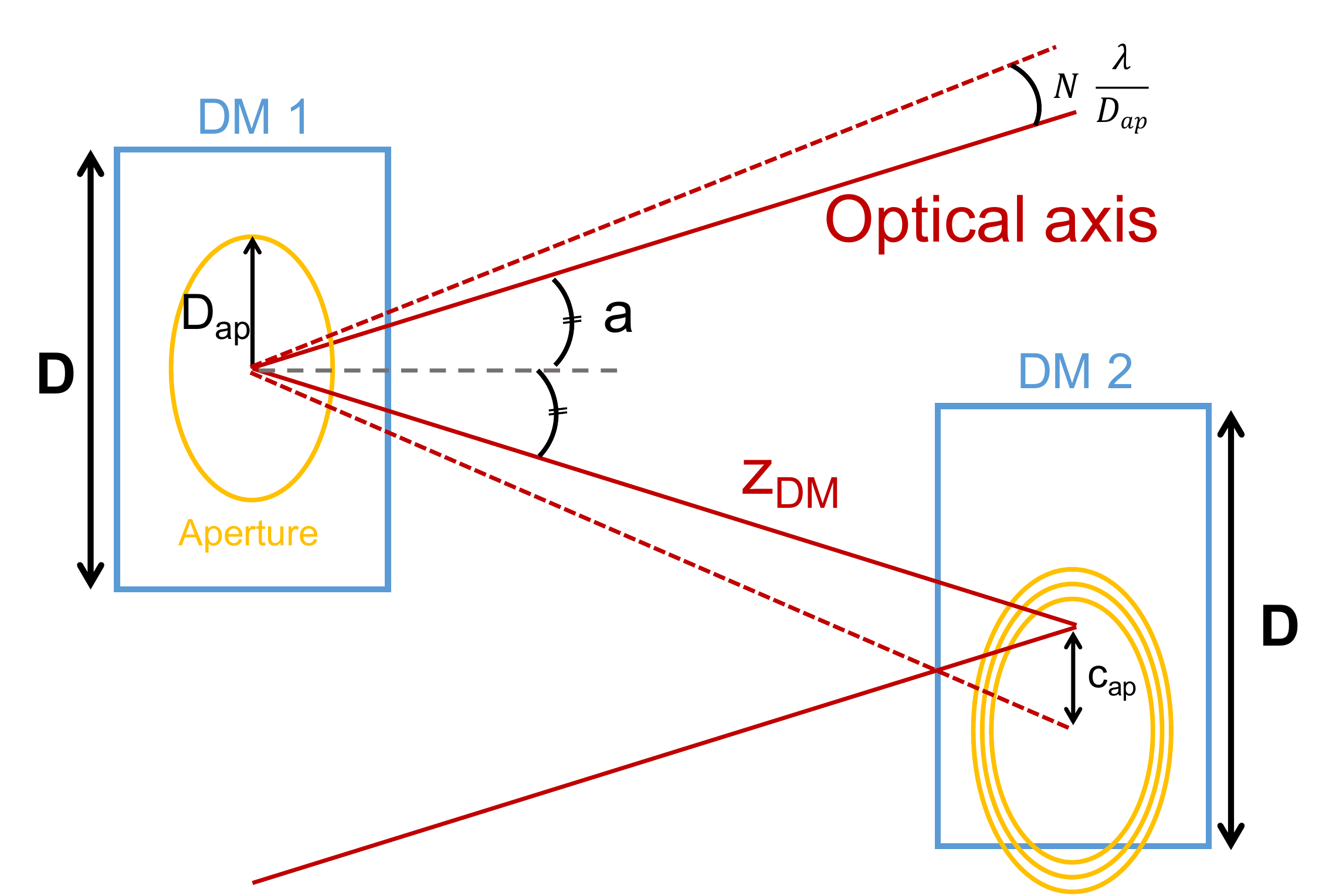}
 \includegraphics[trim= 0.5cm 0.5cm 0.7cm 0.2cm, clip = true,width = .48\textwidth]{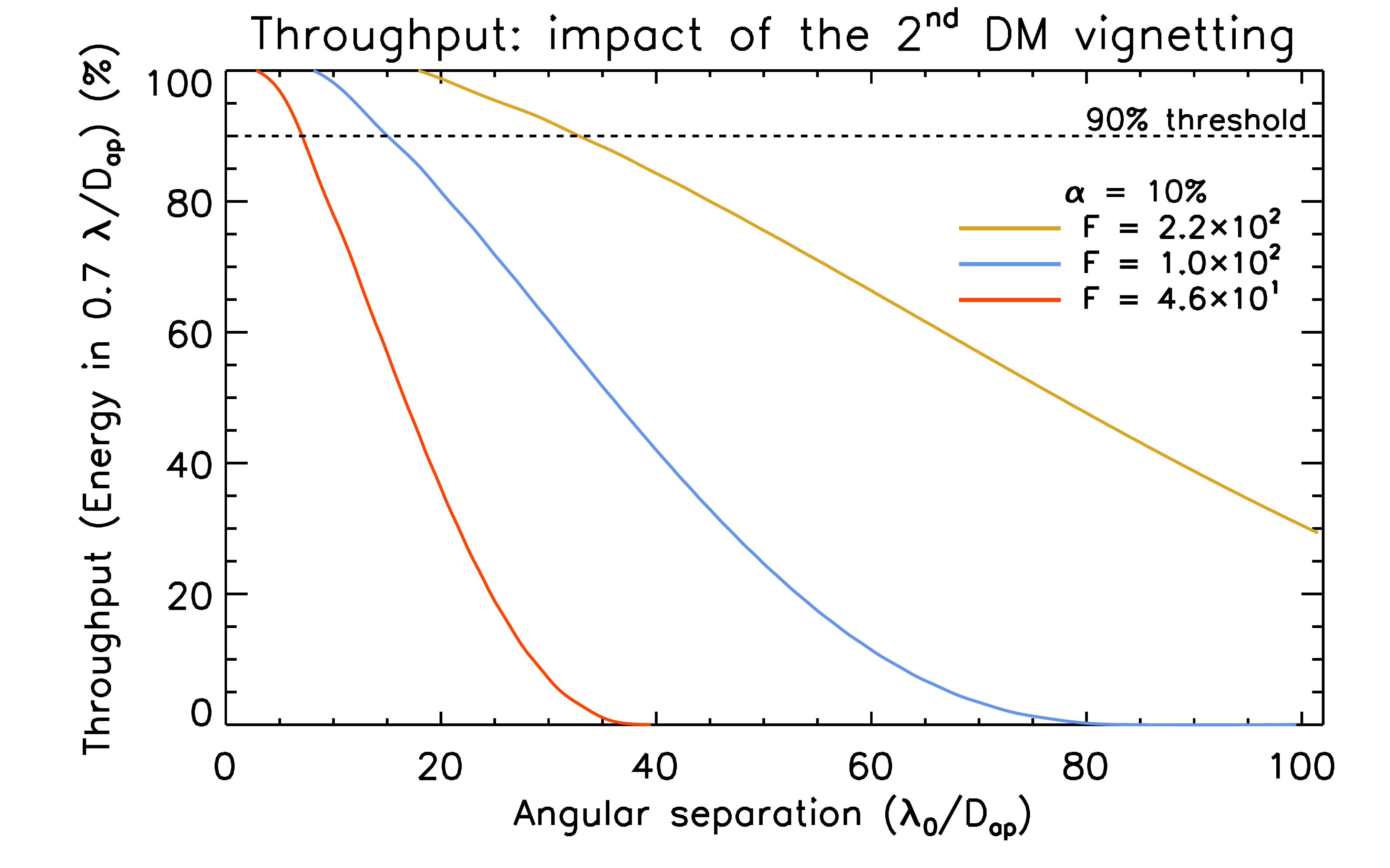}
 \end{center}
\caption[plop]
{\label{fig:schema_throughputlossgeometry_andfuntt_fnum} \textbf{Left:} Geometric schematic showing the effect of the Fresnel number on the throughput of an off-axis planet located at $n\lambda_0/D_{ap}$ of the star. \textbf{Right:} Throughput loss due to the second DM vignetting for 3 Fresnel numbers and $\alpha$ = 10\%.}
\end{figure}

The problem is shown in Fig.~\ref{fig:schema_throughputlossgeometry_andfuntt_fnum} (left). $a$ is the angle between the axis of the system and the DM perpendiculars. We can deduce the distance $c_{ap}$ from the center of the aperture image to the center of the second DM: 
\begin{align*}
    c_{ap} &=  z \sin(a + N\lambda_0/D_{ap}) - z \sin(a)\,\,,\\
    c_{ap} &= z \cos(N\lambda_0/D_{ap})\sin(a) + z \sin(N\lambda_0/D_{ap})\cos(a)- z \sin(a) \,\,,\\
    c_{ap} &\simeq  z N\lambda_0/D_{ap}\cos(a)\,\,,
\end{align*}
If we assume that $N\lambda_0/D_{ap}\ll 1$. Finally, using Eq~\ref{eq:alphadef}, we deduce:
\begin{equation}
c_{ap} \simeq  n\dfrac{(1+\alpha) \lambda_0 z}{D} \cos(a) \,\,.
\end{equation}
To simplify the discussion, we assume that $a = 0$: the axis is perpendicular to both DMs. This case is unrealistic but sets an upper limit to the impact of the effect we are describing here. Assuming flat DMs and an axisymmetric aperture for simplicity, the energy in the beam has a radial symmetry. In that case, half of the diffracted beam is outside of the second DM when $c_{ap}(n) = D/2$, which corresponds to a loss of 50\% of the transmission. We call this separation $n_{\textrm{Tr}50\%}$:

\begin{equation}
\label{eq:n50}
n_{\textrm{Tr}50\%} =  \dfrac{1}{2 (1+\alpha)}\dfrac{D^2}{\lambda_0 z_{DM}} = \dfrac{1}{2 (1+\alpha)}F_{DM} \,\,.
\end{equation}
This analysis shows that this effect is linearly dependent on the Fresnel number and favors large values of $F$. However, this analytic formula is not a practical tool for designing future two DM coronagraphic instruments: a 50\% loss of the off-axis energy transmission, without even considering the impact of the vignetting on the shape of the PSF, is not acceptable.

\subsection{Numerical analysis of the effect of vignetting}
We now use a simple numerical simulation to estimate the loss of off-axis throughput due to the vignetting of the second DM and set a acceptable threshold at 10\%. We simulate two flat DMs for several Fresnel numbers and $\alpha$ cases, from the initial clear aperture to the entrance pupil plane of the coronagraph. We assume that the two DMs are face to face and perpendicular to the axis ($a = 0$). We then simulate the PSF in the next focal plane (with no coronagraphic FPM) and measure the energy in its core (in a radius of 0.7 $\lambda_0/D_{ap}$ around the expected position of the PSF). This two DM system throughput analysis, with flat mirrors and without any coronagraph, does not take into account the throughput of the coronagraph, nor the off-axis throughput loss due to the strokes on the DMs, which is studied independently in the next section. Fig~\ref{fig:schema_throughputlossgeometry_andfuntt_fnum} (right) shows the loss of throughput at as a function of separation for three Fresnel numbers and $\alpha$ = 10\%. It shows that for small F numbers (F = $4.6\times10^{1}$, red curve), the loss of throughput is important (50\% at 20 $\lambda_0/D_{ap}$). However, for other setups (F = $2.2\times10^{2}$, yellow curve), the loss of throughput starts to be important at large separations, which are usually greater than $N_{act}/2$ (OWA set by the DM).

\begin{figure}
\begin{center}
  \includegraphics[trim= 1.8cm 1cm 1.8cm 0.5cm,width = .48\textwidth]{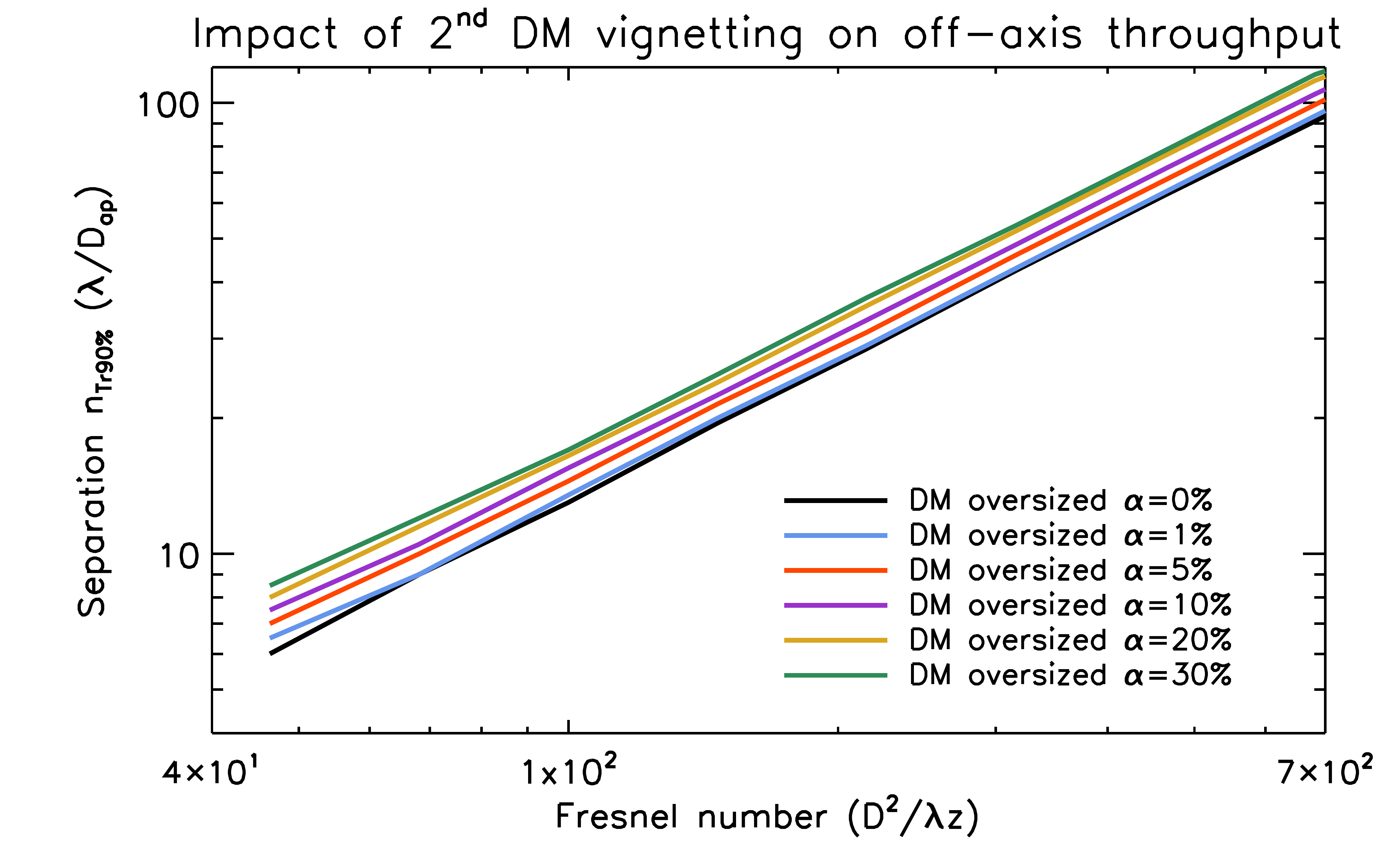}
 \end{center}
\caption[plop]
{\label{fig:oversizing_throughput} These curves represent $n_{\textrm{Th}90\%}$, the separation at which the two DM system throughput is only 90\%, due to the vignetting of the second DM, as function of the Fresnel number for several oversizing cases.}
\end{figure}

To understand the impact of this effect with all DM setups, we used this simple numerical simulation to measure the smallest separation for which the throughput of the two DM system drops under 90\%, $n_{\textrm{Th}90\%}$. The 90\% throughput threshold is represented with a dashed horizontal line in Fig~\ref{fig:throughput_funtt_fnum} and  $n_{\textrm{Th}90\%}$ corresponds to the abscissa where this line intersects the colored throughput curves. We ran this simulation for several F numbers and several over-sizing cases ($\alpha$ from 0 to 30 \%), and plot the results in Fig.~\ref{fig:oversizing_throughput}. Only the results for $n_{\textrm{Th}90\%} < 110 \lambda_0/D_{ap}$ and F $ < 7\times10^{2}$ are plotted.

These curves confirmed the linear trend predicted by Eq.~\ref{eq:n50}. This effect in practice prevents the use of any system with $F < 2\times10^{2}$, corresponding to $n_{\textrm{Th}90\%} < 30 \lambda_0/D_{ap}$, or even further depending on the aimed-for DH size. This effect might have an impact on the maximum OWA that achievable for a given Fresnel number. However, if $n_{\textrm{Th}90\%} \gg  N_{act}/2$ this effect is negligible and the limiting factor for the OWA remains the number of actuators on the DM.

In the rest of this article, we assume a setup where the second DM is surrounded by a non-actuated reflective surface that extends the side length of the second mirror to twice the length of the DM. This allows us to ignore the two DM system vignetting effects on throughput described in this section. All the throughput performance shown in this paper are therefore only due to either the coronagraph or the amount of strokes on the DMs.

\section{Theoretical limitations of two DM correction in the Talbot Regime}
\label{sec:talbot}

In this section, we now specifically explore the limitations to the correction of a simple phase (Sec~\ref{sec:phase_rip}) and amplitude ripple (Sec~\ref{sec:ampl_rip}) of $N$ cycle in the aperture. We specifically assume that we are in the Talbot regime ($F_{DM}\gg N^2$) in this section.

\subsection{Correction of small errors: phase ripple}
\label{sec:phase_rip}

We start by trying to correct a phase ripple in the pupil plane. For a given number $N$, number of cycles inside the aperture of diameter $D_{ap}$, we make the assumption that we can write the field in the pupil plane (z = PP) as:
\begin{equation}  
E_{ap}(\lambda,N,z = PP) = e^{ \dfrac{4 i \pi \lambda_0}{\lambda} \epsilon[N] \cos\left( \dfrac{2 \pi N x}{D_{ap}}\right)} 
\end{equation}
For a small ripple, we have:
\begin{equation}
\label{eq:field_aperture}
E_{ap}(\lambda,N,z = PP) \simeq 1 +  \dfrac{4 i \pi \lambda_0}{\lambda} \epsilon[N] \cos\left( \dfrac{2 \pi N x}{D_{ap}}\right)
\end{equation}
$\epsilon[N]$ is the $N^{th}$ coefficient of the decomposition of the discontinuities in Fourier series. In the general case, it is therefore the amplitude of the electrical field created by this aperture at $N \lambda_0/D_ap$ in the focal plane. In our case, because we only care about the frequency inside the DH we will always have $N \le OWA$.
We call: 
\begin{equation}
X =   \dfrac{2 \pi x}{D_{ap}}
\end{equation}
and we can write our phase ripple as:
\begin{equation}
E_{ap}(\lambda,N) \simeq 1 + \dfrac{4 i \pi \lambda_0}{\lambda} \epsilon[N]\cos(NX)
\end{equation}
That can be corrected with only one DM in pupil plane:
\begin{equation}
E_{DM1}(\lambda,N) = \exp \left[ \dfrac{4 i \pi}{\lambda} \phi_{DM1} \right] = \exp \left[  \dfrac{4 i \pi \sigma_{DM1}\lambda_0  }{\lambda}   \cos(NX) \right]
\end{equation}
First, we assume that the field $A\epsilon[N]$ is small enough so that the strokes introduced by the DM to correct for it still verify $ \sigma_{DM1} \ll  1$.
\begin{equation}
E_{DM1}(\lambda,N) = 1+  \dfrac{4 i \pi \lambda_0  }{\lambda} \sigma_{DM1}  \cos(NX) + \mathcal{O}(\sigma_{DM1}^2)
\end{equation}
With:
\begin{equation}
\label{eq:sigmaphase}  
 \sigma_{DM1} = - \epsilon[N]
\end{equation}
We can correct for this ripple perfectly at all wavelengths and the stroke $\sigma_{DM1}$ on the first DM only depends on the amplitude of the incoming phase ripple.

\subsection{Correction of small errors: amplitude error}
\label{sec:ampl_rip}

We now have only a amplitude ripple:
\begin{equation}
E_{ap}(\lambda,N) \simeq 1 + \epsilon \cos(NX)
\end{equation}
With the first DM, we can introduce phase in the focal plane

\begin{equation}
\label{eq:champDM1}
E_{DM1}(\lambda,N,z = PP) = 1+  \dfrac{4 i \pi \lambda_0  }{\lambda} \sigma_{DM1}  \cos(NX) + \mathcal{O}(\sigma_{DM1}^2)
\end{equation}

We introduce a second DM at a distance $z_{DM}$ and try to measure its influence in the PP. The second DM introduces only phase in the DM2 plane. We also assume $\sigma_{DM2} \ll  1$ in the small aberrations assumption.
\begin{equation}
E_{DM2}(\lambda,N,z = DM2) = 1+  \dfrac{4 i \pi \lambda_0  }{\lambda}  \sigma_{DM2} \cos(NX)
\end{equation}
We use Goodman 2005 (p87-89)\cite{goodman05} description of the Fresnel propagation of an sinusoidal grating (also called Talbot effect) to measure the influence of the periodic phase in the DM2 plane. The field introduced in pupil plane (PP) by the second DM is:
\begin{align*}
E_{DM2}(\lambda,N,z = PP) &= 1+  \dfrac{4 i \pi \lambda_0  }{\lambda}\sigma_{DM2}   \cos(NX)\exp\left( - \dfrac{i \pi \lambda z N^2}{D^2}\right)\, 
\end{align*}
We introduce the Fresnel number at the central wavelength $F_{DM}$ (Eq.\ref{eq:fnumDM}) and we can write the electrical field $E_{DM2}$ as:
\begin{equation}
\label{eq:correct_dm2}
E_{DM2}(\lambda,N,z = PP) = 1+  \dfrac{4 i \pi \sigma_{DM2}\lambda_0  }{\lambda}   \cos(NX)\exp\left( - \dfrac{i \lambda \pi N^2}{ \lambda_0 F_{DM}}\right) 
\end{equation}
We define the Talbot-limited range as the configuration in which $F_{DM}\gg N^2$. In this regime, we have:

\begin{equation}
\label{eq:champDM2}
 E_{DM2}(\lambda,N,z = PP) = 1+  \dfrac{4 i \pi\lambda_0  }{\lambda} \sigma_{DM2}  \cos(NX)\left( 1 - \dfrac{i\lambda \pi N^2}{ \lambda_0 F_{DM}} + \dfrac{1}{2}\left(\dfrac{\pi \lambda N^2}{\lambda_0 F_{DM}}\right)^2 \right)
\end{equation}
The on-axis part is removed by the coronagraph and we only keep the term in $\cos(NX)$. The field in pupil plane (we always have z = PP from now on) is now, separated in real and imaginary part: 

\begin{equation}
\label{eq:field_toto}
E_{Tot}(\lambda,N) = \left[\epsilon[N] + \dfrac{4 \pi^2 N^2}{F_{DM}} \sigma_{DM2} \,\,\,\,\,\, , \,\,\,\,\,\, \dfrac{4 \pi \lambda_0  }{\lambda} (\sigma_{DM1} +\sigma_{DM2}) + \dfrac{2 \pi^3 N^4  }{F_{DM}^2}  \dfrac{\lambda}{\lambda_0} \sigma_{DM2}\right]
\end{equation}
The second DM is correcting for the amplitude:
\begin{equation}
\epsilon[N] + \dfrac{4 \pi^2 N^2}{F_{DM}} \sigma_{DM2} = 0
\end{equation}
The dependence in $\lambda$ is the same for the correcting term and for aberration to correct, therefore all amplitude at every wavelengths is corrected when:
\begin{equation}
\label{eq:sigma_DM2}
\sigma_{DM2} = - \dfrac{\epsilon[N] F_{DM}}{4  \pi^2 N^2}
\end{equation}
Contrarily to the correction with 1 DM (Eq. \ref{eq:sigmaphase}), this shows that the strokes required on the second DM scales linearly with the Fresnel number $F_{DM}$. We now need to correct the phase with the first DM. However, some of the terms we seed to correct do not have the same wavelength dependence as the correction term in $\sigma_{DM1}$. Indeed, we need to find $\sigma_{DM1}$ such as :
\begin{equation}
\label{eq:imagin_part}
 \dfrac{4 \pi \lambda_0  }{\lambda} (\sigma_{DM1} +\sigma_{DM2}) + \dfrac{ 2\pi^3 N^4  }{F_{DM}^2}  \dfrac{\lambda}{\lambda_0} \sigma_{DM2} = 0
\end{equation}
The first DM has to correct for the phase introduced by the second DM (same chromatic dependence, possible at all wavelengths), but also for the phase introduced by the second order of the Talbot effect (different chromatic dependence). When we replace $\sigma_{DM2}$ by its value, Eq.~\ref{eq:imagin_part} reads:
\begin{align*}
\dfrac{4 \pi \lambda_0  }{\lambda} (\sigma_{DM1} +\sigma_{DM2}) - \epsilon[N] \dfrac{ \pi N^2  }{2F_{DM}}\dfrac{\lambda}{\lambda_0}   = 0
\end{align*}
$\sigma_{DM1}$ cannot correct for this field at every wavelengths. We only correct it at the central wavelength $\lambda_0$:
\begin{equation}
\label{eq:sigma_DM1}
\sigma_{DM1}  = \epsilon[N] \left(\dfrac{ N^2}{8F_{DM}} - \dfrac{F_{DM}}{4  \pi^2 N^2} \right)
\end{equation}
This means that, in absolute value, the strokes on the first DM will be less important than on the second DM although close in the Talbot-limited range ($F_{DM}\gg N^2$). We finally write the residual (non corrected) field by replacing $\sigma_{DM1}$ by its value in the field in Eq.~\ref{eq:field_toto}: 

\begin{align*}
E_{res}(\lambda,N) =  \epsilon[N] \dfrac{ \pi N^2  }{2F_{DM}} \left[ \dfrac{\lambda_0  }{\lambda} - \dfrac{\lambda}{\lambda_0} \right]
\end{align*}
This equation assumes that we only can correct in monochromatic light at the central bandwidth and then are left with a residual electrical field when we apply a larger bandwidth. There might be more efficient way to correct in large bandwidth (by minimizing simultaneously at several wavelengths that sampled the bandpass) but they are not studied in this paper. 

The contrast is the residual light level $|E_{res}|^2$ divided by the light that is stripped away by the coronagraph (1 in Eq.~\ref{eq:field_aperture}):
\begin{equation}
C_1(\lambda,N) = \epsilon[N]^2 \dfrac{ \pi^2 N^4  }{4F_{DM}^2} \left[ \dfrac{\lambda_0  }{\lambda} - \dfrac{\lambda}{\lambda_0} \right]^2
\end{equation}
We integrate over the bandwidth and obtain:
\begin{equation}
C_1(N) = \epsilon[N]^2 \dfrac{ \pi^2 N^4  }{4F_{DM}^2} \left(\dfrac{1}{4R^2 - 1} + \dfrac{1}{12R^2} \right)
\end{equation}
where R = $\Delta\lambda/\lambda_0$ is the spectral resolution. We finally integrate over the DH. We can use the following formula for the PSD:
\begin{equation}
\label{eq:formulaPSD}
\epsilon [N] = \dfrac{PSD_0}{N^\beta}\,\,\,\,,\,\,\,\,\, \beta>0
\end{equation}
In that case we can integrate $\epsilon [N]$ for $C_1$, $C_2$ and $C_3$.
\begin{align*}
C_1(N) = \dfrac{ PSD_0^2 \pi^2 N^{4-2\beta}  }{4F_{DM}^2} \left(\dfrac{1}{4R^2 - 1} + \dfrac{1}{12R^2} \right)
\end{align*}
and then integrate over the DH:
\begin{equation}
C_{DH,1} =  \dfrac{PSD_0^2 \pi^2 }{4F_{DM}^2} \left(\dfrac{OWA^{5-2\beta} -IWA^{5-2\beta}}{5 -2\beta } \right) \left(\dfrac{1}{4R^2 - 1} + \dfrac{1}{12R^2} \right) 
\end{equation}
 We don't know the sign of $5-2\beta$ in the general case:
\begin{equation}
C_{DH,1} =  \dfrac{PSD_0^2 \pi^2 }{12} \left(\dfrac{OWA^{5-2\beta} -IWA^{5-2\beta}}{5 -2\beta } \right) \dfrac{1}{R^2} \dfrac{1}{F_{DM}^2}
\end{equation}

However, this formula is only valid if $\sigma_{DM1}\ll 1$, $\sigma_{DM2}\ll 1$. However, sometimes this approximation cannot be done. Indeed Eq.~\ref{eq:sigma_DM2} and \ref{eq:sigma_DM1} shows that at the first order:
\begin{equation}
\label{eq:approxsigma}
\sigma_{DM1}\sim - \sigma_{DM2} \sim  \dfrac{A\epsilon[N] F_{DM}}{4  \pi^2 N^2}
\end{equation}
This is one of the major results of this proceeding: the strokes on the DMs scale with the Fresnel number in the Talbot limited range. This means that for high $F_{DM}$ numbers, the strokes on the DMs will eventually grow too high to use the approximation in Eq.~\ref{eq:champDM1} (even for small aberrations in the aperture).\\

\subsection{Large strokes and frequency folding limitation}
\label{sec:}

In that case, we cannot neglect the second order therm in Eq.~\ref{eq:champDM1}. This term have been called the frequency-folding term in previous work \cite{giveon06}. From Eq.~\ref{eq:champDM1}, we develop to the next term:
\begin{align*}
E_{DM1}(\lambda,N) &= 1+  \dfrac{4 i \pi \lambda_0  }{\lambda} \sigma_{DM1}  \cos(NX) + \dfrac{1}{2}\left(\dfrac{4 i \pi \lambda_0  }{\lambda} \sigma_{DM1}\right)^2  \cos^2(NX) + \mathcal{O}(\left( \dfrac{A\epsilon[N] F_{DM}}{N^2}\right)^3)\\
E_{DM1}(\lambda,N) &= 1 +\dfrac{1}{4}\left(\dfrac{4 i \pi \lambda_0  }{\lambda} \sigma_{DM1}\right)^2+  \dfrac{4 i \pi \lambda_0  }{\lambda} \sigma_{DM1}  \cos(NX) + \dfrac{1}{4}\left(\dfrac{4 i \pi \lambda_0  }{\lambda} \sigma_{DM1}\right)^2  \cos(2NX)+ \mathcal{O}(\left( \dfrac{A\epsilon[N] F_{DM}}{N^2}\right)^3)
\end{align*}
This new term has an impact on the speckle at half-frequency ($f = 1/(2N)$), which is the reason it is called frequency folding. However, if we write this last equation for the double frequency ($f = 2/N$), the frequency folding will introduce an amplitude term in 1/$\lambda^2$ in $\cos(NX)$. This term can be written as:
\begin{align*}
E_{ff,DM1}(\lambda,N) = \dfrac{1}{4}\left(\dfrac{4 i \pi \lambda_0  }{\lambda} \sigma_{DM1}[f = 2/N]\right)^2 \cos(NX)\\
E_{ff,DM1}(\lambda,N) = -(4\pi)^2\left(\dfrac{\lambda_0 }{\lambda}\right)^2\left(\dfrac{\epsilon[N/2] F_{DM}}{N^2}\right)^2 \cos(NX)
\end{align*}
using approximation for $\sigma_{DM1}[f = 2/N]$ in Eq.~\ref{eq:approxsigma}. We also have the same term for the second DM ($E_{ff,DM2} \simeq E_{ff,DM1}$). Once again we remove the on-axis term and only keeps the $\cos(NX)$ terms. The total field at the N frequency now becomes:
\begin{align*}
E_{Tot}(\lambda,N) = \Bigg[\epsilon[N] + \dfrac{4 \pi^2 N^2}{F_{DM}} \sigma_{DM2} + \left[1 - \dfrac{1}{2}\left(\dfrac{\pi \lambda N^2}{\lambda_0 F_{DM}}\right)^2 \right](E_{ff,DM1} +E_{ff,DM2}) ,\\ 
\dfrac{4 \pi \lambda_0  }{\lambda} (\sigma_{DM1} +\sigma_{DM2})+  \dfrac{2 \pi^3 N^4  }{F_{DM}^2}  \dfrac{\lambda}{\lambda_0} \sigma_{DM2} + \dfrac{\lambda \pi N^2}{ \lambda_0 F_{DM}}(E_{ff,DM1} +E_{ff,DM2})\Bigg]
\end{align*}
and we replace $E_{ff,DM1}$ and $E_{ff,DM2}$ by there values
\begin{align*}
E_{Tot}(\lambda,N) = \Bigg[\epsilon[N] + \dfrac{4 \pi^2 N^2}{F_{DM}} \sigma_{DM2}  -  32 \pi^2 \left(\dfrac{\lambda_0 }{\lambda}\right)^2\left(  \dfrac{\epsilon[N/2] F_{DM}}{N^2}\right)^2 + (2\pi)^4 (\epsilon[N/2])^2,\\ 
\dfrac{2 \pi \lambda_0  }{\lambda} (\sigma_{DM1} +\sigma_{DM2}) +  \dfrac{ \pi^3 N^4  }{F_{DM}^2}  \dfrac{\lambda}{\lambda_0} \sigma_{DM2} +  32\pi^3 \dfrac{\lambda_0  }{\lambda} \dfrac{(\epsilon[N/2])^2 F_{DM}}{N^2}\Bigg]
\end{align*}
As in the previous case, we correct for the amplitude term (real part of $E_{Tot}$) with the second DM. The term in $\epsilon[N]$ have been corrected in the first part. The term in $(\epsilon[N/2])^2$ has the same chromatic dependence as the correction term in $\sigma_{DM2}$. It can be corrected at all wavelengths with:
\begin{equation}
\sigma_{DM2,(\epsilon[N/2] )^2} = - 4\pi^2 \dfrac{(\epsilon[N/2])^2 F_{DM}}{ N^2}
\end{equation}
which is negligible compared to the strokes introduced in Eq.~\ref{eq:sigma_DM2}. The real problem comes from the term in $1/\lambda^2$. We can correct it at the central wavelength with 
\begin{equation}
\label{eq:strokeDM2_secondpart}
\sigma_{DM2,1/\lambda^2} = - 8\dfrac{(\epsilon[N/2])^2 F_{DM}^3}{ N^6}
\end{equation}
and the residual field is 
\begin{equation}
E_{res_{ff},amp}= -  32 \pi^2\left(\dfrac{\epsilon[N/2] F_{DM}}{N^2}\right)^2 \left[ 1-\left( \dfrac{\lambda_0 }{\lambda}\right)^2 \right]
\end{equation}
And the contrast is:
\begin{equation}
C_{2}(N,\lambda)=  \left(32 \pi^2\right)^2 \left(\dfrac{\epsilon[N/2] F_{DM}}{N^2}\right)^4 \left[ 1-\left( \dfrac{\lambda_0 }{\lambda}\right)^2 \right]^2
\end{equation}
We integrate over the bandpass:
\begin{equation}
C_{2} (N)=  \left(32 \pi^2\right)^2 \left(\dfrac{\epsilon[N/2] F_{DM}}{N^2}\right)^4 \left[  \dfrac{1}{3 R^2} + \dfrac{1}{80 R^4}\right] 
\end{equation}
We finally use Eq.\ref{eq:formulaPSD} to integrate over the DH:
\begin{equation}
C_{DH,2} =   (32 \pi^2)^2  \left(\dfrac{PSD_0}{2^\beta}\right)^4 \dfrac{ F_{DM}^4}{4\beta+7}  \left[ \dfrac{1}{3R^2} + \dfrac{1}{80 R^4}\right] \left[ IWA^{-4\beta-7} - OWA^{-4\beta-7}\right] 
\end{equation}
In that case, because $-4\beta-7 < 0$, we have $ IWA^{-4\beta-7} \gg OWA^{-4\beta-7}$:
\begin{equation}
C_{DH,2} \sim     \dfrac{1024\pi^4  }{3(4\beta+7)} \left(\dfrac{PSD_0}{2^\beta}\right)^4 \dfrac{F_{DM}^4 }{R^2} * IWA^{-4\beta-7} 
\end{equation}

Depending on the value of $\epsilon$, and $F/N^2$, this term is higher or not than the initial contrast $C_{DH,1}$. However, whatever is the initial amplitude, at large enough $F_{DM}$ we have necessarily $C_{DH,2} \gg C_{DH,1}$.\\
Now, just like previously, the strokes we introduced on the second DM in Eq.~\ref{eq:strokeDM2_secondpart} must be corrected in the phase part (imaginary part of $E_{Tot}$) with the first DM. All the terms that have the $\lambda_0/\lambda$ dependence can be corrected at all wavelengths with strokes negligible compare to Eq.~\ref{eq:sigma_DM1}. The real problem comes from the correction of the $\lambda/\lambda_0$ term, that we can correct at the central wavelength with: 
\begin{equation}
\sigma_{DM1,\lambda/\lambda_0} = 4\pi^2 \dfrac{(\epsilon[N/2])^2 F_{DM}}{N^2}
\end{equation}
and the resulting field is 
\begin{equation}
E_{3}(N,\lambda)= 16 \pi^3 (\epsilon[N/2])^2 \dfrac{F_{DM} }{N^2} \left[ \dfrac{\lambda_0  }{\lambda} - \dfrac{\lambda}{\lambda_0} \right]
\end{equation}
And the contrast is:
\begin{equation}
C_{3}(N,\lambda)=  256\pi^6  (\epsilon[N/2])^4 \left(\dfrac{F_{DM} }{N^2}\right)^2 \left[ \dfrac{\lambda_0  }{\lambda} - \dfrac{\lambda}{\lambda_0} \right]^2
\end{equation}
after integration on the bandpass is:
\begin{equation}
C_{3}(N)=  256\pi^6  \epsilon[N/2]^4 \dfrac{F_{DM}^2}{N^4} \left(\dfrac{1}{4R^2 - 1} + \dfrac{1}{12R^2} \right) 
\end{equation}
We integrate over the DH, which gives us:
\begin{equation}
C_{DH,3}= 256\pi^6  \left(\dfrac{PSD_0}{2^\beta}\right)^4  \dfrac{F_{DM}^2}{4\beta+3} \left(\dfrac{1}{4R^2 - 1} + \dfrac{1}{12R^2} \right)  \left[ IWA^{-4\beta-3} - OWA^{-4\beta-3}\right] 
\end{equation}
In that case, because $-4\beta-3 < 0$, we have $ IWA^{-4\beta-3} \gg OWA^{-4\beta-3}$:
\begin{equation}
\label{eq:Cres_pha}
C_{DH,3} \sim  \dfrac{256 \pi^6}{3*(4\beta+3)}\left(\dfrac{PSD_0}{2^\beta}\right)^4  \dfrac{F_{DM}^2 }{R^2 }IWA^{-4\beta-3} 
\end{equation}

It is hard to compare $C_{DH,1}$, $C_{DH,2}$ and $C_{DH,3}$ in the general case. However, there is a few things we can safely say: 
\begin{itemize}
    \item in the Talbot limited regime, the strokes increase with the Frenel number and with the amount of aberrations, at some point, the correction will be limited by this effect.
    \item before these strokes grows too high, we are in the case of appendix A.1 and the contrast in the DH goes in $C_{DH,3} \sim 1/F_{DM}^2$. This regime is especially valid for the small aberrations case ($\epsilon^2 \ll A^2 \epsilon^4$).
    \item However, for larger and larger $F_{DM}$ numbers, the strokes groes higher and eventually the contrast in the DH goes in  $C_{DH,3} \sim F_{DM}^2$ and then in  $C_{DH,3} \sim F_{DM}^4$.
    \item the wavelength dependence in the Talbot regime is the same for everyone of these contrast regime and is in $\sim 1/R^2$. This dependence was also fund in \cite{shaklan06} for small phase aberrations. We verified this prediction with an end-to-end simulation (results not shown in this proceeding).
\end{itemize}

\subsection{End-to-end simulation with the ACAD-OSM method and the WFIRST aperture}

\begin{figure}
\begin{center}
 \includegraphics[ trim= 1.5cm 1cm 1.0cm 0.5cm, clip = true, width = .48\textwidth]{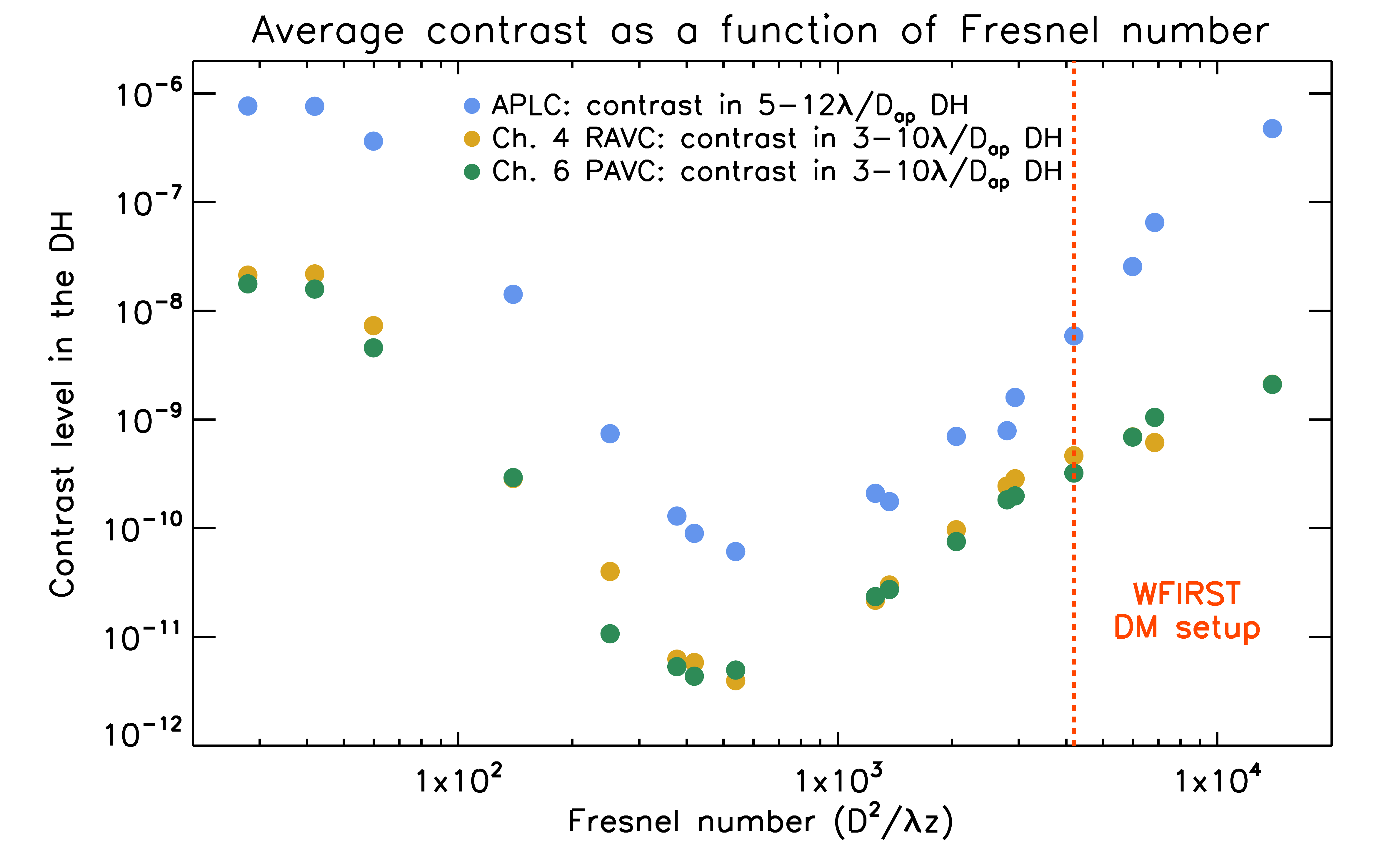}
 \includegraphics[trim= 2.5cm 1cm 1.0cm 0.5cm, clip = true, width = .48\textwidth]{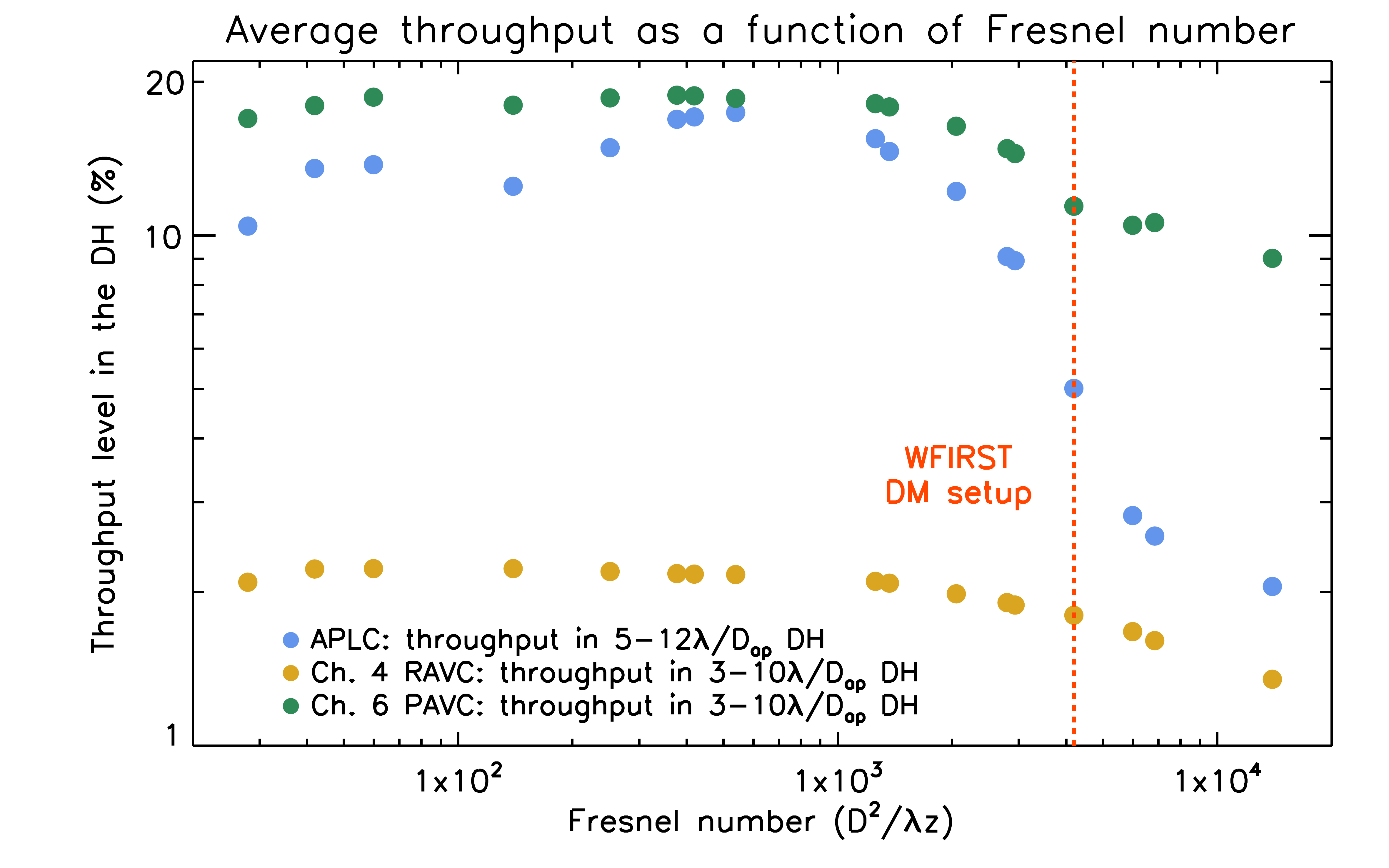}
 \end{center}
\caption[plop]
{\label{fig:perf_contrast_fresnel} Performance of the correction in contrast (\textbf{left}) and in throughput (\textbf{right}) as a function of the Fresnel number, for the WFIRST aperture, a 10\% BW around 550 nm, and for 48 actuators. The vertical red dashed line indicates the actual WFIRST DM setup.}
\end{figure}

In this section, we show the results of the end-to-end simulation realized with the ACAD-OSM method with the WFIRST aperture\cite{spergel15} (shown in Fig~\ref{fig:optical_schema_2DM}). We made 17 simulations at different Fresnel number, with 3 different coronagraph. These coronagraph were optimized to correct for the central obstruction of this aperture and the ACAD-OSM method was only used to correct for struts of the secondary. The 3 coronagraph were: a charge 6 polynomial apodized vortex coronagraph (PAVC, see paper \#10400-27, Fogarty et al. in these proceedings and \cite{fogarty17}), a charge 4 ring apodized vortex coronagraph (RAVC) \cite{mawet13} and an apodize pupil Lyot coronagraph (APLC) \cite{ndiaye16}. The simulation were realized with a 10\% bandwidth. We always use the same number of actuators ($N_{act} = 48$). The OWA is 10  $\lambda_0/D_{ap}$ for the vortexes and $12 \lambda_0/D_{ap}$ for the APLC. 

Fig.\ref{fig:perf_contrast_fresnel} shows the performance in contrast level (left) and in throughput (right), as a function of the Fresnel number for these three coronagraphs. The first interesting aspect is that points with different distances and DM sizes but with similar Fresnel numbers gives similar performance in contrast and throughput, proving that the performance is not dependent on $z$ or $D$ independently but only on the Fresnel number as expected. 

For $F>5\times10^{2}$, we are in the Talbot effect-limited regime ($OWA^2 = 100$ for the vortexes and 144 for the APLC). We fit a line to the increasing slope of this curve and found contrast degrades as $F^{2}$ as expected from the $C_{DH,2}$ contrast residual. We do not explain the limitation at small Fresnel number in this proceedings but in an upcoming article\cite{mazoyer17b} 

For throughput, we see that the throughput performance is almost flat until $F \sim 10^{3}$ but then decreases quickly at higher Fresnel numbers. However, one has to remember that we assumed that the surface outside of the second DM is reflective, although not actuated. We showed in Section~\ref{sec:vignetting}, that without this assumption, we would have a important loss of off-axis throughput for $F < 2\times10^{2}$ due to the 2 DM system vignetting.

\section{Is there really an OWA limit associated with optics outside of the pupil?}

\begin{figure}
\begin{center}
 \includegraphics[width = .48\textwidth]{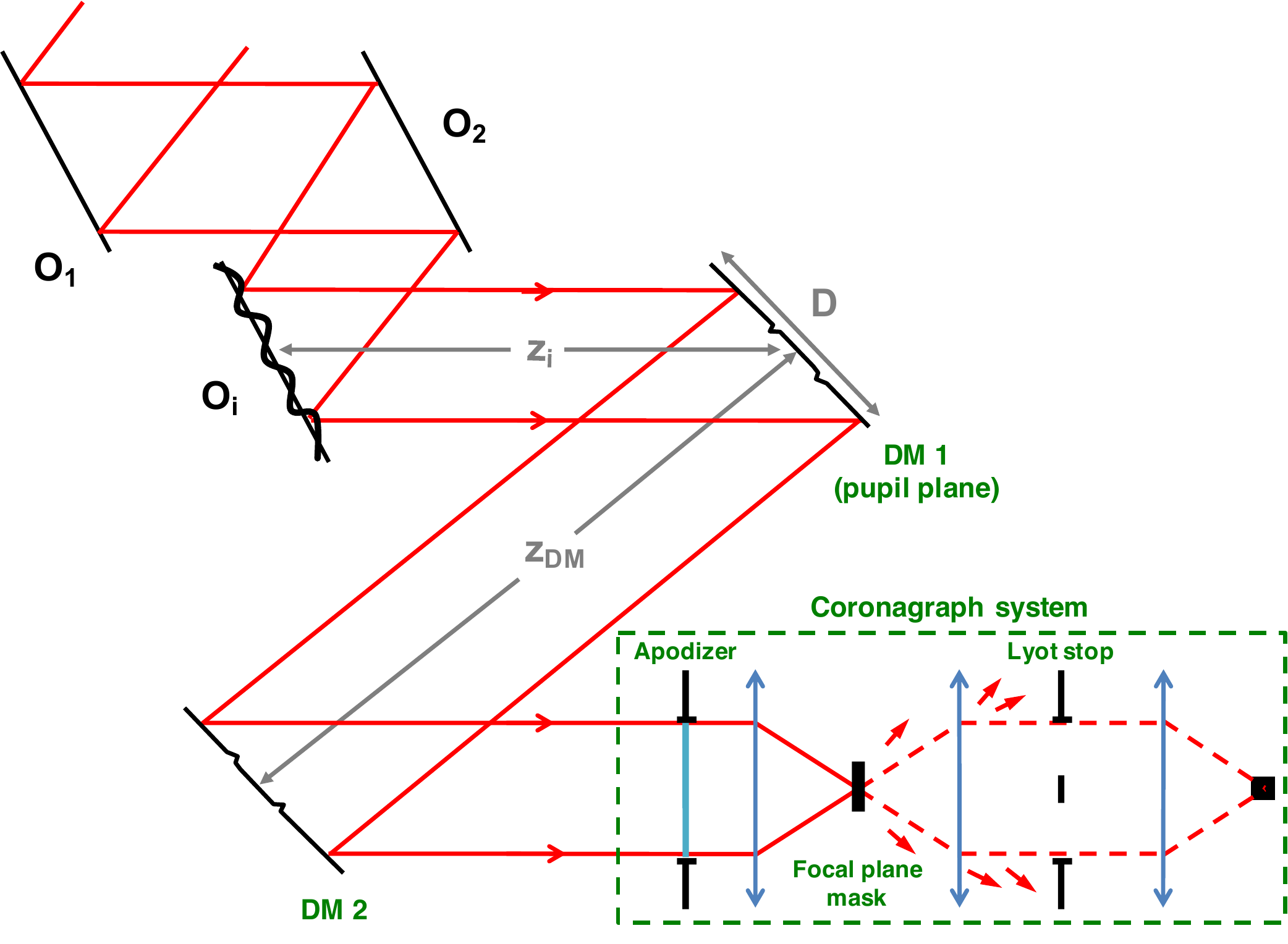}
 \end{center}
\caption[plop]
{\label{fig:optical_schema_upfront_optics} Schematic representation of a two DM system and a coronagraph with upfront optics at a distance $z_i$.}
\end{figure}

We now change the scope of this study, using the same Talbot formalism. Fig~\ref{fig:optical_schema_upfront_optics} shows a schematic of a realistic case for a telescopes. Before reaching the DMs and the coronagraph, the beam encounter several optics $O_1$, $O_2$, ..., $O_i$... These optics do not have perfect surface quality, and the phase aberrations they introduce can limit the contrast in practice in the focal plane. We assume a phase ripple, not on the pupil plane, but on a i-th optics $O_i$, at a distance $z_i$ from the pupil plane:

\begin{equation}
E_{i}(\lambda,N, z = O_i) \simeq 1 + \dfrac{4 i \pi \lambda_0}{\lambda} \epsilon_i[N]\cos(NX)
\end{equation}
In this part, we are not studying the frequency folding effects, and we will always assume $\epsilon_i[N] = \epsilon_i$. When the ripple reaches the pupil plane, we have:
\begin{equation}
\label{eq:before_talbot_Assuption}
E_{i}(\lambda,N, z = PP) = 1+  \dfrac{4 i \pi \lambda_0}{\lambda} \epsilon_i   \cos(NX)\exp\left( - \dfrac{i \lambda \pi N^2}{ \lambda_0 F_i}\right) 
\end{equation}
where $F_i$ is the Fresnel number of the couple of optics (DM1, $O_i$):
\begin{equation}
F_i = \dfrac{D^2}{\lambda_0 z_i}
\end{equation}
In the previous part and in \cite{shaklan06}, the assumption that we were in the  Talbot regime ($F_i \gg N^2$) was made. Here, we do not make this assumption: in the general case, $F_i \sim N^2$, and therefore, we have:
\begin{align*}
E_{i}(\lambda,N) &= 1+  \dfrac{4 i \pi \lambda_0}{\lambda} \epsilon_i   \left[\cos\left( \dfrac{\lambda \pi N^2}{ \lambda_0 F_i}\right) + i\sin\left( \dfrac{\lambda \pi N^2}{ \lambda_0 F_i}\right)\right]\cos(NX)
\end{align*}
As in the previous case, we assume that the coronagraph takes care of the on-axis components (the $'1'$) and we only correct for the $\cos(NX)$ part. We see now that a simple ripple outside of the pupil plane introduces phase and amplitude aberrations in the pupil plane and that both these aberrations are dependent on the Fresnel number of this optics $F_i$. In the next sections, we analyze how these frequencies are limiting the performance in contrast, first in the case of 1 DM, then for 2 DMs. 

\subsection{One DM correction}

\subsubsection{Correction of the phase with one DM}

We have one DM in pupil plane and we seek to create a ripple with this DM at the same spatial frequency $N$ that cancels the imaginary part of the field in the pupil.
\begin{eqnarray}
 0 & =& 4 \pi \frac{\lambda_0}{\lambda}\cos(NX) ( \epsilon_i \cos\left( \dfrac{\lambda \pi N^2}{ \lambda_0 F_i}\right)  - \sigma_{DM1})
\end{eqnarray}

We see that the field and the correction term do not have the same wavelength variations, so we cannot correct it on the whole bandwidth. We only correct it at $\lambda = \lambda_0$ with:
 \begin{eqnarray}
  \sigma_{DM1}& =& \epsilon_i \cos\left(\dfrac{ \pi N^2}{F_i}\right)
\end{eqnarray}
Once again, we note that there might be a ``less chromatic'' wavefront control algorithm where we try to correct for several wavelength at the same time. We do not explore this in this proceeding. The residual field is:

\begin{equation}  
 E_{res, 1DM, \phi}(\lambda,N)  =   \frac{\epsilon_i (4 \pi \lambda_0)}{\lambda^2}\left(\cos\left(\dfrac{\lambda \pi N^2}{ \lambda_0 F_i}\right) - \cos\left(\dfrac{\pi N^2}{  F_i}\right) \right)
\end{equation}

Then the residual contrast $C_{1DM, \phi}$ of that speckle can be seen as the integrated squared intensity of the Fourier coefficient in the pupil:

\begin{eqnarray}  
C_{1DM, \phi}(i,N) & = & \epsilon_i^2 \frac{\lambda_0}{\Delta \lambda} (4 \pi)^2 \bigintss_{\lambda_0 - \Delta \lambda /2}^{\lambda_0 - \Delta \lambda /2} \frac{\lambda_0}{\lambda^2}\left(\cos\left(\dfrac{\lambda \pi N^2}{ \lambda_0 F_i}\right) - \cos\left(\dfrac{\pi N^2}{  F_i}\right) \right)^2 d\lambda\\
C_{1DM, \phi}(i,N) & = & \epsilon_i^2 R (4 \pi)^2 I_1(\lambda_0, \Delta \lambda, z_i, z_{DM}, D, N)
\end{eqnarray}
where, $I_1$ is the integral. We do not try to solve analytically this integral and only compute it. For a given contrast goal $C_g$ and a given bandwidth, we have, at every frequencies $N$:
\begin{equation}  
\epsilon_i(i,N)  = \sqrt{\dfrac{C_g}{(4 \pi)^2 R I_1}}
\end{equation}
This is the requirement on the quality of the optics outside of the optical plane to achieve the contrast $C_g$

\subsubsection{Correction of the phase and amplitude with one DM}

We now seek to create a ripple with the DM at the same spatial frequency $N$ (but shifted by $\pi/2$ for a half dark hole) which cancels the real part of the field in the pupil at $\lambda = \lambda_0$ 
\begin{eqnarray}
 && 0  = 4 \pi \frac{\lambda_0}{\lambda} \left( \epsilon_i \sin \left( \dfrac{\lambda \pi N^2}{ \lambda_0 F_i}\right)\cos(NX)   - \sigma_{DM1} \cos(NX) \right)
\end{eqnarray}
We correct at the central wavelength with:
\begin{eqnarray}
\sigma_{DM1} = \epsilon_i \sin\left( \dfrac{ \pi N^2}{ F_i}\right)
\end{eqnarray}

Then the residual contrast $C_{1DM, a}$ is:
 
\begin{eqnarray}  
C_{1DM, a}(i,N) &=& \epsilon_i^2 \frac{\lambda_0}{\Delta \lambda} (4 \pi )^2 \bigintss_{\lambda_0 - \Delta \lambda /2}^{\lambda_0 - \Delta \lambda /2} \frac{\lambda_0}{\lambda^2}\left(\sin \left( \dfrac{\lambda \pi N^2}{ \lambda_0 F_i}\right) - \sin \left( \dfrac{ \pi N^2}{ F_i}\right) \right)^2 d\lambda \\
C_{1DM, a}(i,N) &=& \epsilon_i^2 R (4 \pi)^2 I_2(\lambda_0, \Delta \lambda, z_i, z_{DM}, D, N)
\end{eqnarray}
Once again, we can compute the integral and invert to find a requirement on optical quality for every frequency $N$ as a function of contrast goal, bandwidth $\Delta \lambda $, central wavelength $ \lambda_0 $.

\subsubsection{Numerical simulation}
Even though we do not solve analytically the integrals, we prove in simulation that once again, the requirement depends only of $F_i$ and R and not of $D$, $\lambda_0$ and $z_i$ independently.
\begin{eqnarray}  
I_1(\lambda_0, \Delta \lambda, z_i, D, N) &=& I_1(N, R, F_i) \\
I_2(\lambda_0, \Delta \lambda, z_i, D, N) &=& I_2(N, R, F_i) 
\end{eqnarray}
Finally, we also find that there is a dependence with R of the requirements (the requirements are twice as hard at 20\% than at 10\%).

\begin{figure}
\begin{center}
 \includegraphics[ width = .45\textwidth]{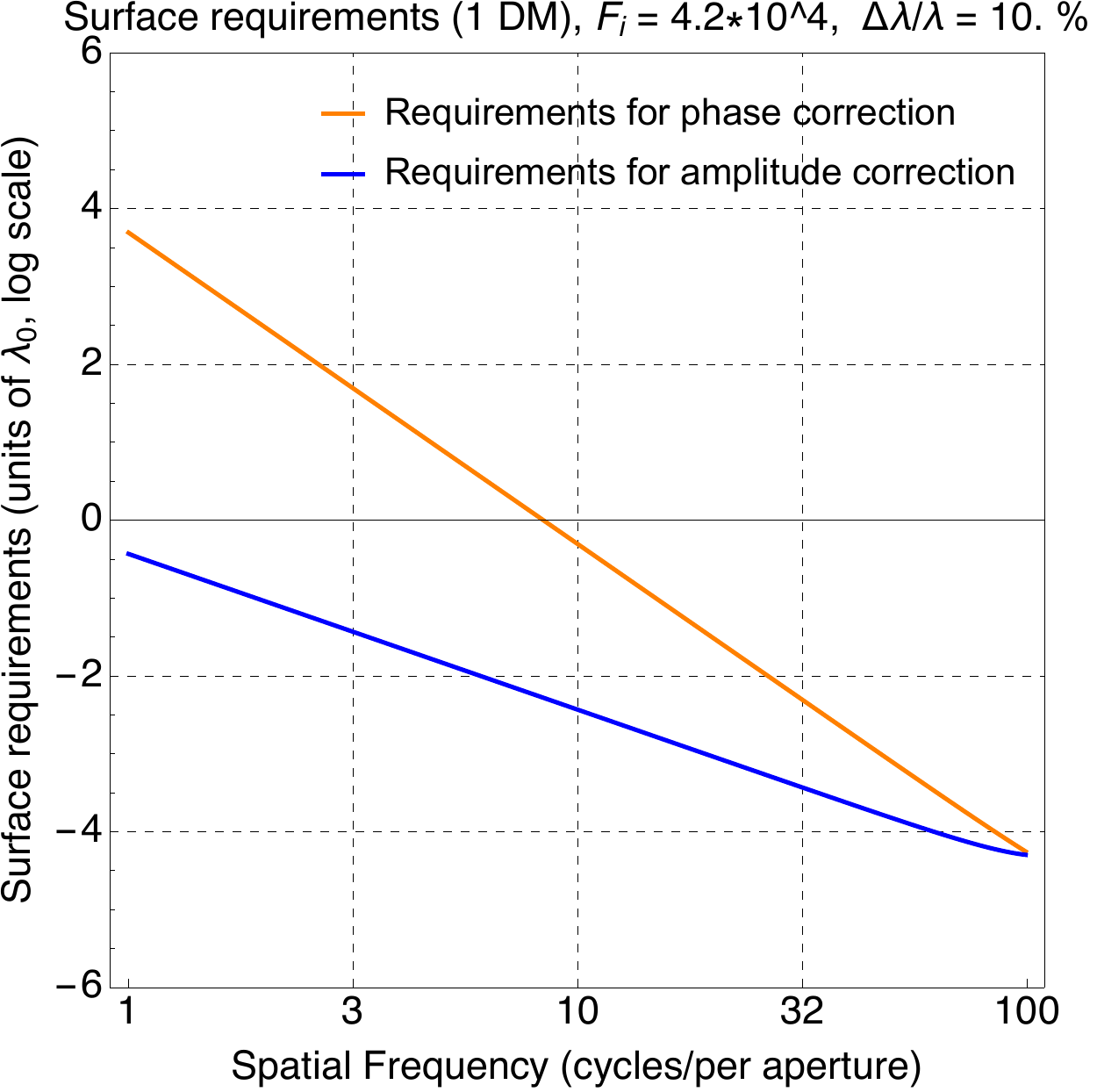}
\includegraphics[ width = .45\textwidth]{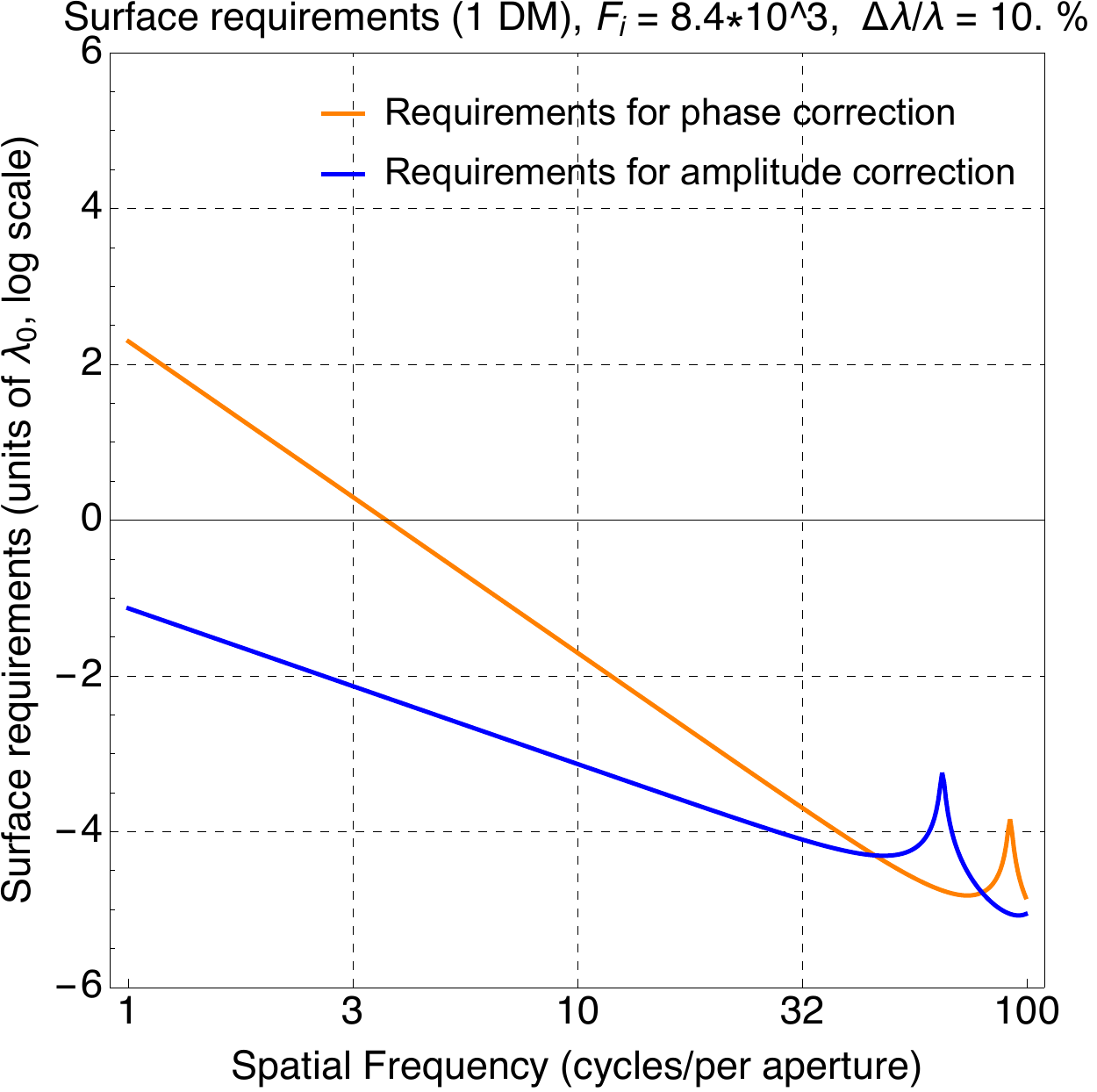}\\
\includegraphics[ width = .45\textwidth]{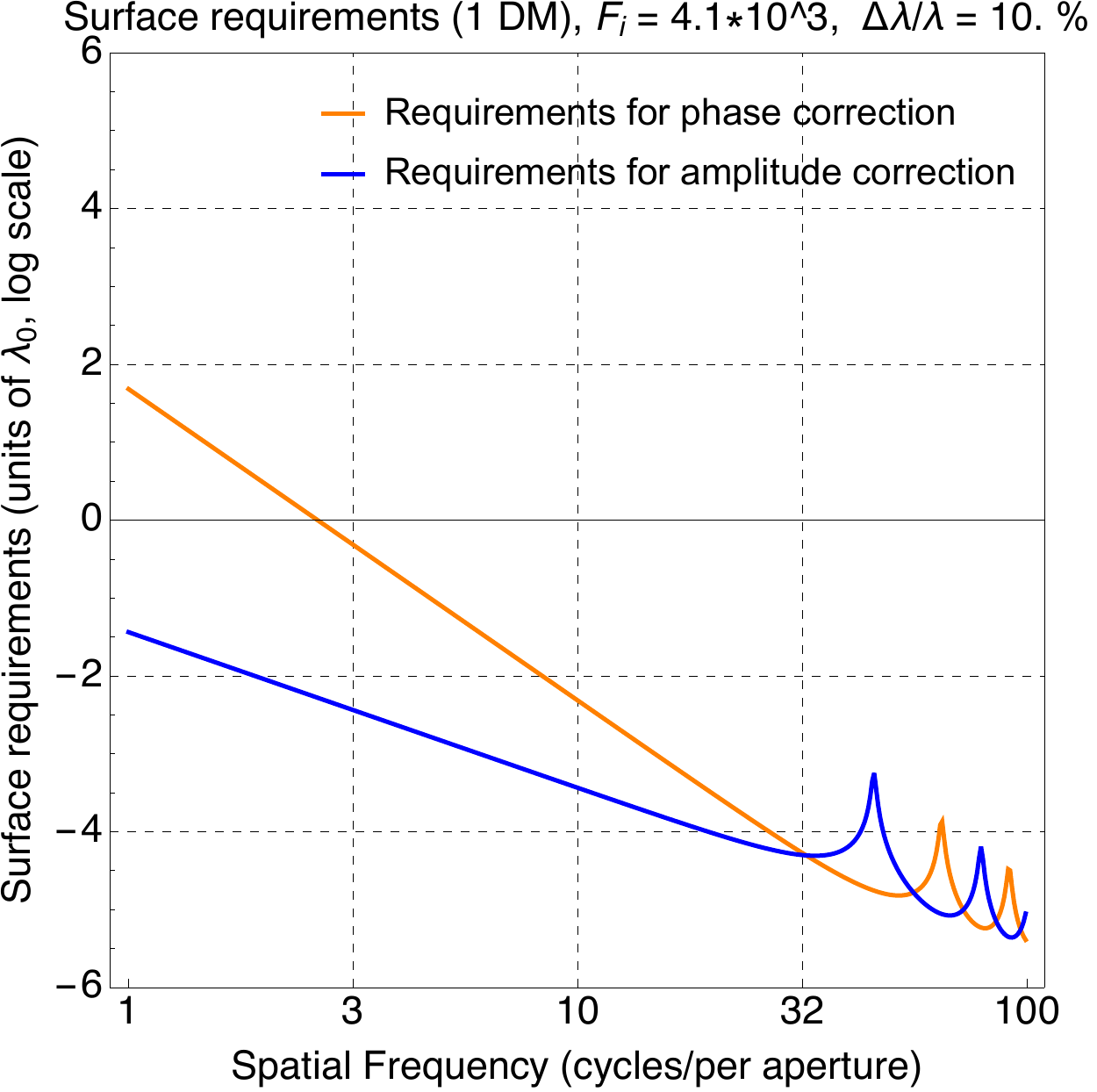}
\includegraphics[ width = .45\textwidth]{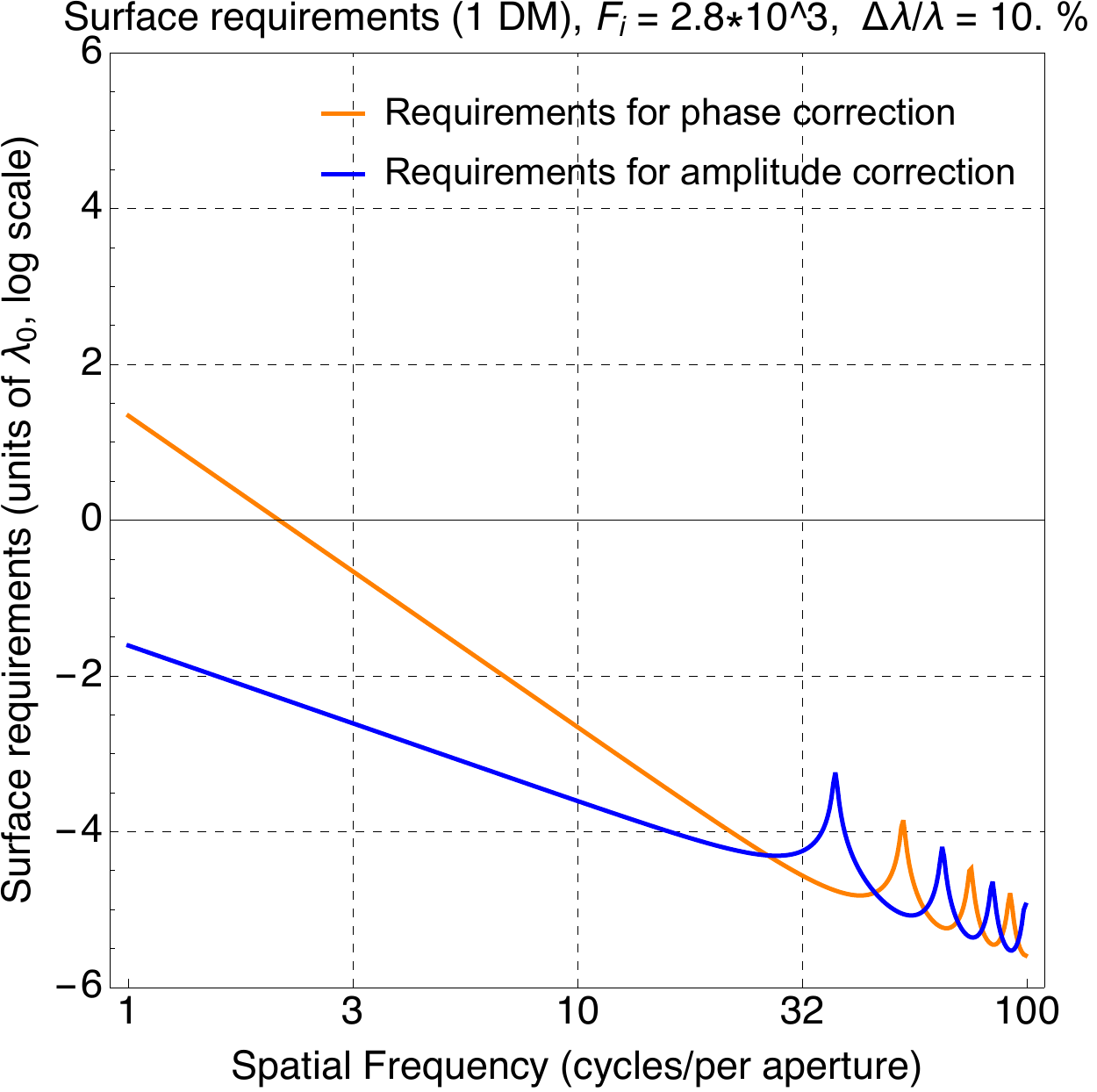}
 \end{center}
\caption[plop]
{\label{fig:1dmrequirements} Curves of requirements for one DM and a contrast goal $C_g = 10^{-10}$, as a function of the number $N$ of cycles in the aperture of the ripple, for $\lambda_0/\Delta \lambda = 10\%$,  and for 4 different $F_i$.}
\end{figure}

We plot in Fig.~\ref{fig:1dmrequirements} the curves of requirements for one DM correction as a function of the number $N$ of cycles in the aperture of the ripple, for $\lambda_0/\Delta \lambda = 10\%$, a contrast goal of $C_g = 10^{-10}$, and for 4 different $F_i$. On the left part of these plots, we have $F_I \gg N$ and it shows a linear trends in the log-log diagram that we would probably could have had if we made this assumption in Eq.~\ref{eq:before_talbot_Assuption}. On the right, we see the waves comings from the sinus and cosinus. 

We see that in practice, the Fresnel number chosen of the optics will set the maximal OWA that one can achieve with this setup. In practice, it is better to chose optics with a large Fresnel number (\textit{i.e.} that are close to the pupil plane).

All the optics $O_1$, $O_2$, ... $O_i$ upfront the two DM and coronagraph system will create their own constraints only depending on their own Fresnel number. The limitation on the OWA will actually be set by the worst optic, at very small Fresnel number (the furthest from the DM1 plane/pupil plane, bottom right in Figure~\ref{fig:1dmrequirements}).

\subsection{Two DM correction}
We now introduce a second DM at a distance $z_{DM}$ of the pupil plane and try to correct for the phase and amplitude terms

\subsubsection{Correction of the amplitude with two DMs}

We seek to create a ripple with DM2 which, when projected at DM1, cancels the real part of the field in the pupil. The field introduced by the second DM in the pupil plane depends on the Fresnel number of the two DM system $F_{DM}$, described in Eq~\ref{eq:correct_dm2}:
\begin{equation}
 0  = 4 \pi \frac{\lambda_0}{\lambda} \cos(NX) \left( \epsilon_i \sin \left(\dfrac{\lambda \pi N^2}{ \lambda_0 F_i}\right)  - \sigma_{DM_2} \sin\left(\dfrac{\lambda \pi N^2}{ \lambda_0 F_{DM}}\right)   \right)
\end{equation}
We correct at the central wavelength $\lambda = \lambda_0$  only:
\begin{equation}
\sigma_{DM_2} = \epsilon_i \frac{\sin(\dfrac{ \pi N^2}{  F_i})}{\sin(\dfrac{ \pi N^2}{  F_{DM}})}
\end{equation}

Then the contrast $C_{2DM, a}$ is:

\begin{eqnarray}  
C_{2DM, a}(i,N)  &=& \epsilon_i^2 \frac{\lambda_0}{\Delta \lambda} (4 \pi )^2 \bigintss_{\lambda_0 - \Delta \lambda /2}^{\lambda_0 - \Delta \lambda /2} \frac{\lambda_0}{\lambda^2}\left(\sin \left(\dfrac{\lambda \pi N^2}{ \lambda_0 F_i}\right) - \frac{\sin(\dfrac{ \pi N^2}{  F_i})}{\sin(\dfrac{ \pi N^2}{  F_{DM}})}\sin\left(\dfrac{\lambda \pi N^2}{ \lambda_0 F_{DM}}\right)  \right)^2 d\lambda\\
C_{2DM, a}(i,N) &=& \epsilon_i^2 R (4 \pi)^2 I_3(\lambda_0, \Delta \lambda, z_i, z_{DM}, D, N)
\end{eqnarray}
Where $I_3$ is the integral that we can compute. This equation can be inverted into a requirement as a function of contrast and other parameters. We know have to correct for the phase aberrations with the first DM.

\subsubsection{Correction of the phase with two DMs}

\begin{figure}
\begin{center}
 \includegraphics[ width = .45\textwidth]{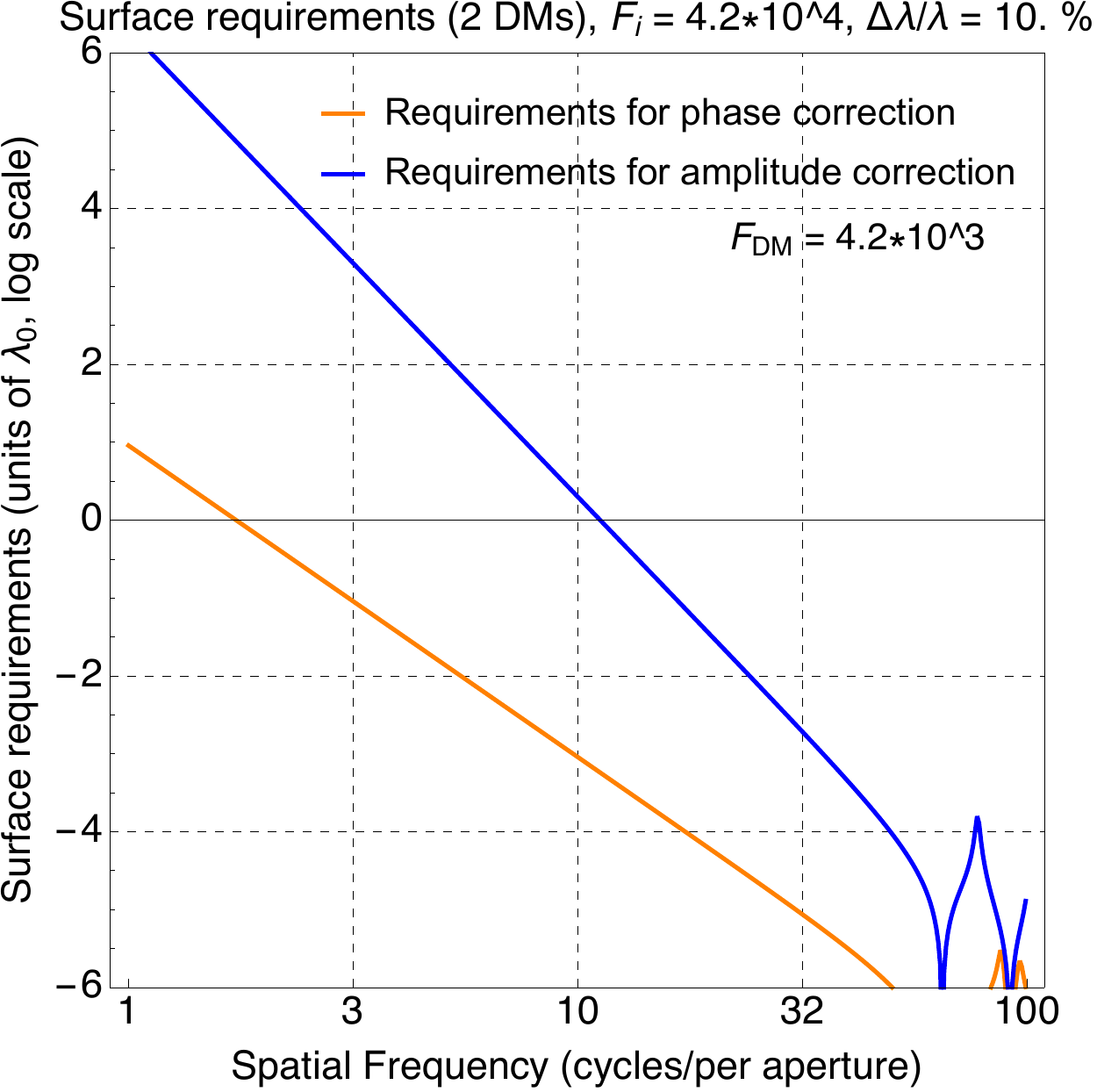}
\includegraphics[ width = .45\textwidth]{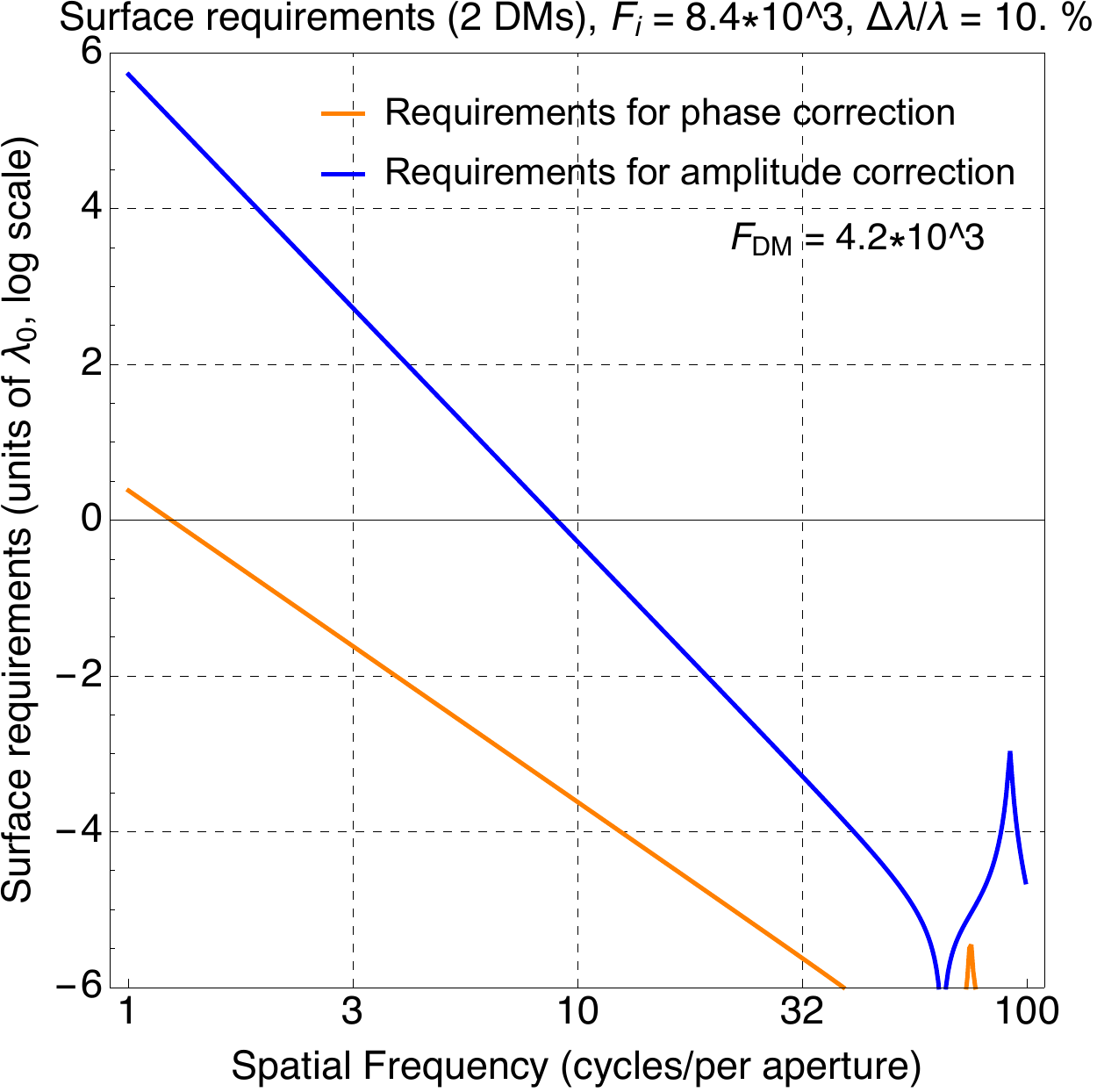}\\
\includegraphics[ width = .45\textwidth]{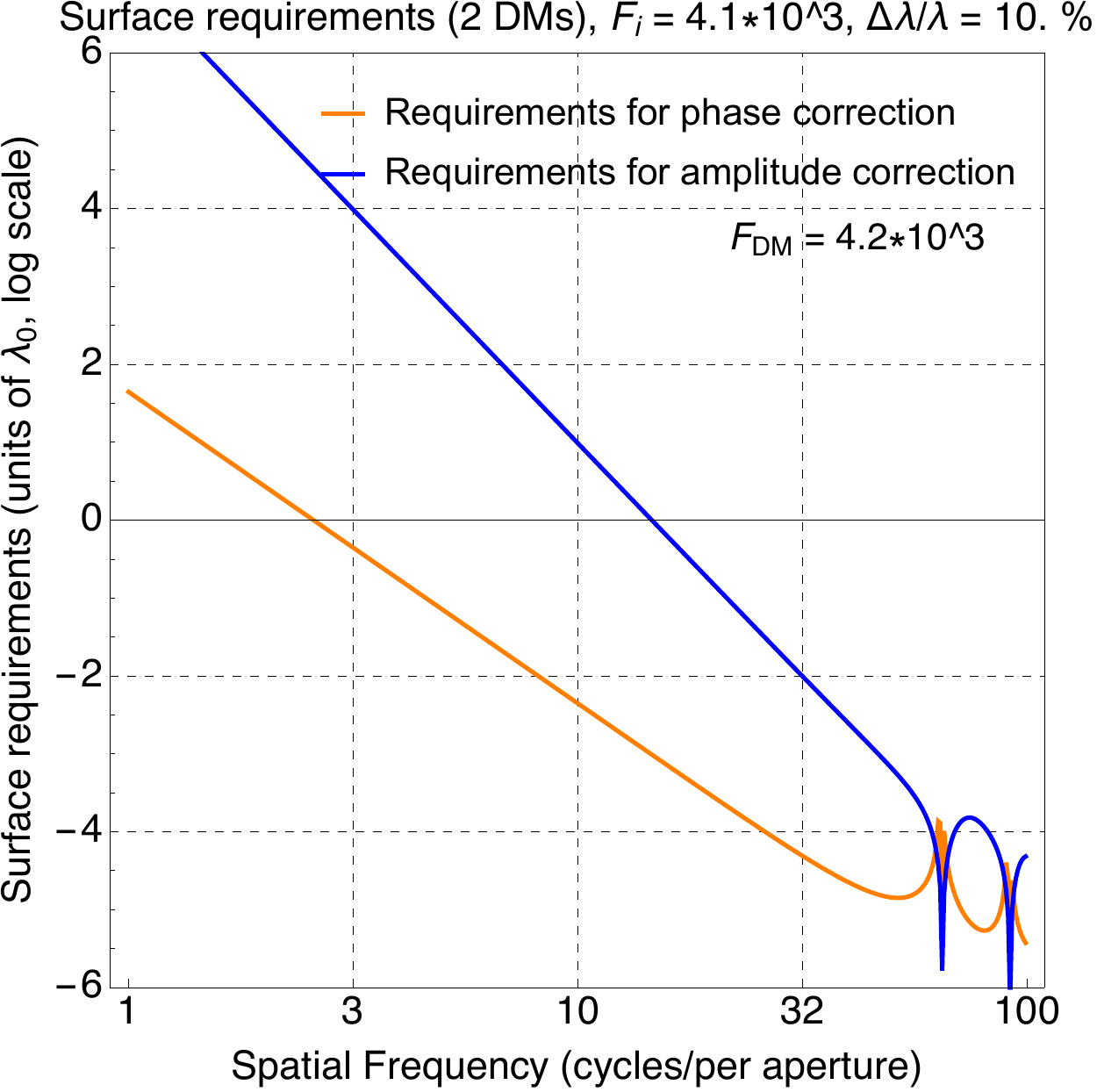}
\includegraphics[ width = .45\textwidth]{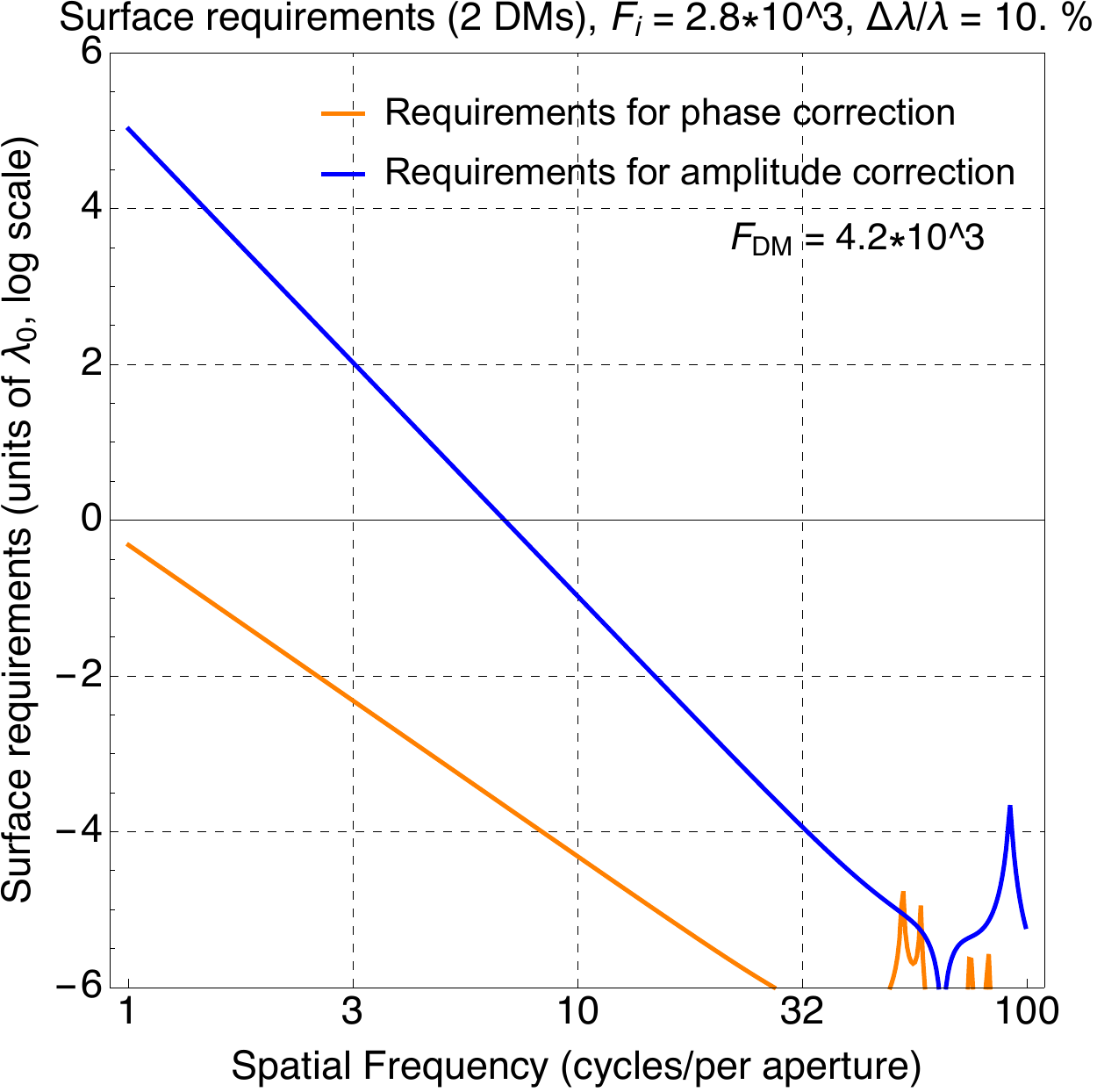}
 \end{center}
\caption[plop]
{\label{fig:2dmrequirements} Curves of requirements for two DM and a contrast goal $C_g = 10^{-10}$, as a function of the number $N$ of cycles in the aperture of the ripple, for $\lambda_0/\Delta \lambda = 10\%$,  and for 4 different $F_i$. Here, we have $F_{DM} = 4.2 \times 10^3$. }
\end{figure}
The DM 1 now has to correct phase introducing a ripple at this frequency. There are two contributions of phase in this system: the ripple of the optic $O_i$ that creates phase in the pupil plane, but also the phase introduce by DM2 in the pupil plane when it corrects the amplitude:
\begin{equation}
 0  = 4 \pi \frac{\lambda_0}{\lambda} \cos(NX) \left( \epsilon_i \cos \left(\dfrac{\lambda \pi N^2}{ \lambda_0 F_i}\right)   - \epsilon_i \frac{\sin(\dfrac{ \pi N^2}{  F_i})}{\sin(\dfrac{ \pi N^2}{  F_{DM}})}\cos\left(\dfrac{\lambda \pi N^2}{ \lambda_0 F_{DM}}\right)  - \sigma_{DM_1}   \right) \nonumber 
\end{equation}
We correct at central wavelength, with:
\begin{equation}
\sigma_{DM_1} = \epsilon_i  \left( \cos \left(\dfrac{ \pi N^2}{  F_i}\right)   -  \frac{\sin(\dfrac{ \pi N^2}{  F_i})}{\sin(\dfrac{ \pi N^2}{  F_{DM}})}\cos \left( \dfrac{ \pi N^2}{  F_{DM}}\right)  \right)
\end{equation}

The contrast $C_{2DM, \phi}$ is:

\begin{eqnarray}  
C_{2DM, \phi}(i,N)   &=&\epsilon_i^2 \frac{\lambda_0}{\Delta \lambda} (4 \pi )^2 \bigintss_{\lambda_0 - \Delta \lambda /2}^{\lambda_0 - \Delta \lambda /2} \frac{\lambda_0}{\lambda^2}\left[ \cos \left(\dfrac{\lambda \pi N^2}{ \lambda_0 F_i}\right)  -  \frac{\sin(\dfrac{ \pi N^2}{  F_i})}{\sin(\dfrac{ \pi N^2}{  F_{DM}})}\cos\left(\dfrac{\lambda \pi N^2}{ \lambda_0 F_{DM}}\right)   -  \right.\\
 &&\left. \left( \cos \left(\dfrac{ \pi N^2}{  F_i}\right)   -  \frac{\sin(\dfrac{ \pi N^2}{  F_i})}{\sin(\dfrac{ \pi N^2}{  F_{DM}})}\cos \left( \dfrac{\pi N^2}{  F_{DM}}\right)  \right) \right]^2 d\lambda\\
 C_{2DM, \phi}(i,N)  &=& \epsilon_i^2 R (4 \pi)^2 I_4(\lambda_0, \Delta \lambda, z_i, z_{DM}, D, N)
\end{eqnarray}
Where $I_4$ is the integral that we can compute. This equation can be inverted into a requirement as a function of contrast and other parameters.

\subsubsection{Numerical simulation}
Even though we do not solve analytically the integrals, we prove in simulation that once again, the requirement depends only of $F_i$ and R and not of $D$, $\lambda_0$ and $z_i$ independently.
\begin{eqnarray}  
I_3(\lambda_0, \Delta \lambda, z_i, z_{DM}, D, N) &=& I_3(N, R, F_i, F_{DM}) \\
I_4(\lambda_0, \Delta \lambda, z_i, z_{DM}, D, N) &=& I_4(N, R, F_i, F_{DM}) 
\end{eqnarray}

We plot in Fig.~\ref{fig:2dmrequirements} the curves of requirements for two DM correction as a function of the number $N$ of cycles in the aperture of the ripple, for $\lambda_0/\Delta \lambda = 10\%$, a contrast goal of $C_g = 10^{-10}$, and for 4 different $F_i$. The $F_{DM}$ is the one currently chosen for the WFIRST mission (D = 48mm, $z_{DM}$ = 1m, $\lambda_0 $ = 550 nm,  $F_{DM} = 4.2 \times 10^3$).

We see that in practice, the Fresnel number chosen of the optics will set the maximal OWA that one can achieve with this setup. In practice, it is better to chose optics with a large Fresnel number (\textit{i.e.} optics that are close to the pupil plane/plane of the first DM) or close to the Fresnel number of the DMs (\textit{i.e.} optics that are close to plane of the second DM). 

All the optics $O_1$, $O_2$, ... $O_i$ upfront the two DMs and coronagraph system will create their own constraints only depending on their own Fresnel number relative to the Fresnel number of the 2 DM system. The limitation on the OWA will actually be set by the worst optics: either at very small Fresnel number (far from both the DM planes, bottom right in Figure~\ref{fig:2dmrequirements}), or in between the plane of DM1 ($F_i$ = 0) and the plane of DM 2: $F_i = F_{DM}$ (top right in Figure~\ref{fig:2dmrequirements}).

\section{Conclusion}

In this proceedings, we show several tools to understand and predict the limitations of active system for high contrast wavefront control. In the first section, we have shown that the second DM vignetting can be a problem at very small Fresnel number, because it limits the off-axis throughput achievable by the system. In the second part, we saw that in the Talbot regime ($F_{DM} \gg N^2$), the strokes needed to correct for amplitude aberrations increase linearly with $F_{DM}$. This constraint limits both the contrast and the throughput of the system at large Fresnel number. Finally in the last section, we analyze the effects of several optics located upfront of the pupil plane and and put requirements on their surface quality.

The aim of these techniques is not to replace end-to-end simulation but to use fast theoretical tools to efficiently design coronagraphic instruments. These predicted effects still have to be confirmed in end-to-end simulations.

\acknowledgements
This material is based upon work partially carried out under subcontract \#1496556 with the Jet Propulsion Laboratory funded by NASA and administered by the California Institute of Technology.

\bibliography{biblio_spie_17}   
\bibliographystyle{spiebib}   

\end{document}